# Network Throughput Optimization via Error Correcting Codes

Ratko V. Tomic[*]
Infinetics Technologies, Inc.

A new network construction method is presented for building of scalable, high throughput, low latency networks. The method is based on the **exact equivalence** discovered between the problem of maximizing network throughput (measured as bisection bandwidth) for a large class of practically interesting Cayley graphs and the problem of maximizing codeword distance for linear error correcting codes. Since the latter problem belongs to a more mature research field with large collections of optimal solutions available, a simple translation recipe is provided for converting the existent optimal error correcting codes into optimal throughput networks. The resulting networks, called here **Long Hop** networks, require 1.5-5 times fewer switches, 2-6 times fewer internal cables and 1.2-2 times fewer 'average hops' than the best presently known networks for the same number of ports provided and the same total throughput. These advantage ratios increase with the network size and switch radix.

Independently interesting byproduct of the discovered equivalence is an efficient $O(n \cdot \log(n))$ algorithm based on Walsh-Hadamard transform for computing **exact bisections** of this class of Cayley graphs (this is NP complete problem for general graphs).

---

[*] Chief Scientist at Infinetics Technologies, Inc., http://www.infinetics.com
 Email: rvtomic@infinetics.com

# 1. Problem Motivation

Rapid proliferation of large Data Center and storage networks in recent years has spurred great deal of interest from industry and academia in optimization of network topologies [1]-[12]. The urgency of these efforts is further motivated by the inefficiencies and costs of the presently deployed large Data Center networks which are largely based on non-scalable tree topology.

There are two main types of network topologies proposed as scalable alternatives to the non-scalable tree topology of the conventional Data Center:

- **Fat Tree** (**FT**) (syn. folded Clos) based networks, a class of "indirect networks"
- **Hypercube-like** (**HC**) networks, a class of "direct networks" using Cartesian product construction recipe. This class includes plain hypercube variants (BCube, MDCube), Folded Hypercube (FC), Flattened Butterfly (FB), HyperX (HX), hyper-mesh, hyper-torus, … etc.

While the HC networks are overall the more economical of the two types, providing the same capacity for random traffic as FT with fewer switches and fewer cables, the FT is more economical on the worst case traffic, specifically on the task of routing the worst case 1-1 pairs permutation.

The **Long Hop** networks (**LH**) described in this paper are above this dichotomy by being simultaneously the most optimal for the common random traffic and for the worst case traffic. The LH optimality is result of the new approach to network construction which is fundamentally different from the techniques used to construct the leading alternatives in the literature. Namely, while the alternative techniques build the network via simple, mechanically repetitive design patterns ('*cookie cutter*' networks) which are not directly related to the network performance metrics such as throughput, the LH networks are constructed via an exact combinatorial optimization of the target metrics, the network throughput.

Although there have been some previous attempts to optimize the network throughput directly, such as the "entangled networks" described in [2] and [12], these techniques sought to optimize general random networks. Since such optimization is computationally intractable for general graphs (it is an NP-complete problem), the computations of both, the network performance and the search for its improvements, are by necessity very approximate (simulated annealing) and still, they become prohibitively expensive as the network size $n$ increases beyond few thousand nodes. For example, the largest computed size in [12] had $n$=2000 nodes, while present DC networks already require scales into hundreds of thousands of nodes. Further, since the resulting approximate solutions have variable node degree and random connectivity, appearing to a network technician as massive, incoherent tangles of wires without any pattern or logic, the "entangled networks" are in practice virtually impossible to wire and troubleshoot. Finally, the node degree irregularity and the complete lack of symmetry of such networks compound their impracticality due to complicated, resource hungry routing algorithms and forwarding tables.

In contrast, the LH construction optimizes the highly symmetrical and, from practical perspective, the most desirable subset of general networks, Cayley graphs [11]. As result of its more focused and more careful identification of the target domain, the LH networks are optimal regarding throughput and latency within that domain, practical to compute and discover, simple and economical to wire and troubleshoot and highly efficient in routing and forwarding resources (they are "self-routing" networks).

# 2. Mathematical Tools and Notation

Since the Long Hop construction intertwines results from several fields of mathematics, computer science, physics and programming not commonly brought together, for convenience of specialists in individual fields, this section provides harmonized notation along the with brief summaries of key concepts and results needed later.

## A. Notation and Terms

- $\lfloor a \rfloor$      floor($a$): the largest integer $\leq a$
- $\mathbb{V}_n$      $n$-dimensional vector space (over some implicit field $\mathbf{F}_q$)
- $\mathbb{S}(k,n,q)$      $k$-dimensional subspace of $\mathbb{V}_n$ (linear span) over field $\mathbf{F}_q$
- $\langle x|y \rangle$      scalar (dot) product of real vectors $x$ and $y$: $\langle x|y \rangle \equiv \sum_{i=1}^{n} x_i\, y_i$
- $\|x\|$      norm (length) of vector $x$: $\|x\| \equiv \sqrt[2]{\langle x|x \rangle}$
- $a..b$      integer sequence $a, a+1, \ldots, b$ for some integers $a \leq b$
- $\{x: \mathbf{E}(x)\}$      set of elements $x$ for which Boolean expression $\mathbf{E}(x)$ is true
- $\min_E\{set\}$      minimum element of a {set} under condition E; analogously for max{set}
- $a \% b$      "$a$ mod $b$" or "$a$ modulo $b$" (remainder in integer division $a/b$)
- bitwise      operation on bit strings done separately in each bit position
- $\tilde{a}$ or $\overline{a}$      NOT $a$ (bitwise complement, toggles each bit of $a$)
- $a \,\&\, b$      bitwise AND (bitwise $a \cdot b$)
- $a \,|\, b$      bitwise OR (bitwise $a + b - a \cdot b$)
- $a \wedge b$      XOR, exclusive OR (bitwise: $(a+b)$ mod 2, also $a + b - 2 \cdot a \cdot b$)
- $a \oplus b$      modular addition in ring $(\mathbf{Z}_q)^d$: component-wise $(a+b)$ mod $q$
- $a \ominus b$      synonym for $a \oplus (-b)$; for $q=2$: $a \ominus b \Leftrightarrow a \oplus b \Leftrightarrow a \wedge b$ (bitwise XOR)
- $\mathbb{V} = \mathbb{V}_1 \oplus \mathbb{V}_2$      Vector space $\mathbb{V}$ is direct sum of vector spaces $\mathbb{V}_1$ and $\mathbb{V}_2$ '
- $A \circ B = B \circ A$      Objects (matrices, group elements, etc.) **commute** for operation '∘'
- $[\mathbf{E}]$      Iverson bracket ($\mathbf{E}$ is a Boolean expression): $\mathbf{E}$ true (false) $\Rightarrow [\mathbf{E}] \equiv 1\,(0)$
- $\delta_{i,j}$      Kronecker delta: $\delta_{i,j} \equiv [i{=}j]$ i.e. $\delta_{i,j}$ is 1 if $i=j$ and 0 if $i \neq j$
- $\delta_i$      Dirac integer delta: $\delta_i \equiv \delta_{i,0}$ i.e. $\delta_i$ is 1 if $i=0$ and 0 if $i \neq 0$
- $\mathbf{B} = \mathbf{A}^T \equiv \tilde{\mathbf{A}}$      matrix $\mathbf{B}$ is a transpose of matrix $\mathbf{A}$ i.e. elements $\mathbf{B}_{i,j} = \mathbf{A}_{j,i}$
- $\mathbf{A} \otimes \mathbf{B}$      Kronecker product of matrices $\mathbf{A}$ and $\mathbf{B}$
- $\mathbf{A}^{\otimes n}$      Kronecker $n$-th power of matrix $\mathbf{A}$: $\mathbf{A}^{\otimes n} \equiv \mathbf{A} \otimes \mathbf{A} \otimes \cdots \otimes \mathbf{A}$ ($n$ times)
- $\mathbf{A} \times \mathbf{B}$      Cartesian product of sets or groups $\mathbf{A}$ and $\mathbf{B}$
- $\mathbf{A}^{\times n}$      Cartesian $n$-th power of a set or group $\mathbf{A}$
- $\mathbf{C}(n,k)$      Binomial coefficient $\mathbf{C}(n,k) \equiv n!/[k!(n-k)!] = \binom{n}{k}$

**Binary expansion** of a *d*-bit integer $X \Leftrightarrow X = \sum_{\mu=0}^{d-1} x_\mu 2^\mu$ where $x_\mu$ is the "*μ*-th bit of **X**" (bits $x_\mu$ have values 0 or 1). **Bit-string form** of the binary expansion of integer **X** is denoted as: $X = x_{d-1}\ldots x_1 x_0$.

**Parity** of a *d*-bit integer $X = x_{d-1}\ldots x_1 x_0$ is: $\mathbb{P}(X) \equiv (x_0+x_1+\ldots+x_{d-1}) \bmod 2 = x_0 \wedge x_1 \wedge \ldots \wedge x_{d-1}$.

**Hamming weight** $\langle X \rangle$ or $\Delta(X)$ of *n*-tuple $X \equiv x_1 x_2 \ldots x_n$, where $x_i \in [0,q)$, is the number of non-zero symbols in **X**. **Hamming distance** $\Delta(X,Y)$ between *n*-tuples **X** and **Y** is the number of positions *i* where $x_i \neq y_i$. For vectors **X** and **Y** this is equivalent to $\Delta(X,Y) = \langle X - Y \rangle \equiv \Delta(X - Y)$ i.e. to Hamming weight of (**X-Y**). For binary strings this yields $\Delta(X,Y) = \langle X \wedge Y \rangle$ i.e. the Hamming weight of $X \wedge Y$.

**Lee distance** is $\Lambda(X, Y) \equiv \sum_{i=1}^{n} \min\{|x_i - y_i|, q - |x_i - y_i|\}$. **Lee weight** is: $\Lambda(X) \equiv \Lambda(X, 0)$.

**Binary intervals** (or binary tiles) are intervals of size $2^k$ (for $k = 1, 2, \ldots$) such that each "tile" of size $2^k$ starts on an integer multiple of $2^k$ e.g. $[m \cdot 2^k, (m+1) \cdot 2^k)$ for any integer *m* are "binary intervals" of size $2^k$.

**Cyclic group** $Z_n$: set of integers $\{0, 1, \ldots n-1\}$ with integer addition modulo *n* as the group operation. Note that $Z_2$ group operation is equivalent to a single bit XOR operation ($1 \wedge 0 = 0 \wedge 1 = 1$, $0 \wedge 0 = 1 \wedge 1 = 0$). The same symbol $Z_n$ is also used for commutative ring with integer additions and multiplication performed mod *n*.

**Product group** $Z_q^d \equiv Z_q \times Z_q \times \cdots \times Z_q$ (*d* times): extension of $Z_q$ into a *d*-tuple. As with $Z_n$, $Z_q^d$ also denotes a commutative ring in which the $Z_q$ operations (integer +,* mod *q*) are done component-wise.

Finite **Dyadic group** $D^d$ of order $n = 2^d$ is abelian group consisting of all *d*-bit integers $0..n-1$ using bitwise XOR (^) as the group operation. Notes: (i) for $n = 2^d$ and $d > 2 \Rightarrow Z_n \neq D^d$; (ii) $D^d$ is an instance of $Z_2^d$.

```
Y^X 0 1 2 3 4 5 6 7 8 9 A B C D E F
0:  0 1 2 3 4 5 6 7 8 9 A B C D E F :0
1:  1 0 3 2 5 4 7 6 9 8 B A D C F E :1
2:  2 3 0 1 6 7 4 5 A B 8 9 E F C D :2
3:  3 2 1 0 7 6 5 4 B A 9 8 F E D C :3
4:  4 5 6 7 0 1 2 3 C D E F 8 9 A B :4
5:  5 4 7 6 1 0 3 2 D C F E 9 8 B A :5
6:  6 7 4 5 2 3 0 1 E F C D A B 8 9 :6
7:  7 6 5 4 3 2 1 0 F E D C B A 9 8 :7
8:  8 9 A B C D E F 0 1 2 3 4 5 6 7 :8
9:  9 8 B A D C F E 1 0 3 2 5 4 7 6 :9
A:  A B 8 9 E F C D 2 3 0 1 6 7 4 5 :A
B:  B A 9 8 F E D C 3 2 1 0 7 6 5 4 :B
C:  C D E F 8 9 A B 4 5 6 7 0 1 2 3 :C
D:  D C F E 9 8 B A 5 4 7 6 1 0 3 2 :D
E:  E F C D A B 8 9 6 7 4 5 2 3 0 1 :E
F:  F E D C B A 9 8 7 6 5 4 3 2 1 0 :F
    0 1 2 3 4 5 6 7 8 9 A B C D E F
```

**Fig. 2.1**

Fig. 2.1 illustrates the group operation table for group $D^4$ with $n = 2^4 = 16$ elements 0, 1, 2, … F (all numbers are in base 16). Table entry in row Y and column X is the result of bitwise X^Y operation.

## B. Matrices and Vectors in Dirac Notation

Dirac notation (also called "bra-ket" notation, [13]) is a mnemonic notation which encapsulates common matrix operations and properties in a streamlined, visually intuitive form.

**Matrix** [$A_{r,c}$] (also: [**A**] or just **A**) is a rectangular table with $r$ rows and $c$ columns of "matrix elements". An element on $i$-th row and $j$-th column of a matrix [**A**] is denoted as [$A$]$_{i,j}$. Identity matrix $n \times n$ is denoted as $I_n$ or **I**. Matrices with $r = 1$ or $c = 1$, **row** or **column vectors**, are denoted as follows:

$$\text{Row vector (bra):} \quad \langle X| \equiv (x_1 \ x_2 \cdots x_n) \qquad \text{Column vector (ket):} \quad |Y\rangle \equiv \begin{pmatrix} y_1 \\ y_2 \\ \vdots \\ y_n \end{pmatrix}$$

$$\text{Inner (scalar) product:} \quad \langle X|Y\rangle \equiv (x_1 \ x_2 \cdots x_n) \cdot \begin{pmatrix} y_1 \\ y_2 \\ \vdots \\ y_n \end{pmatrix} \equiv \sum_{i=1}^{n} x_i \ y_i = \text{"number"}$$

$$\text{Outer product:} \quad |Y\rangle \langle X| \equiv \begin{pmatrix} y_1 \\ y_2 \\ \vdots \\ y_r \end{pmatrix} (x_1 \ x_2 \cdots x_c) \equiv \begin{pmatrix} y_1 x_1 & y_1 x_2 & \ldots & y_1 x_c \\ y_2 x_1 & y_2 x_2 & \ldots & y_2 x_c \\ \ldots & \ldots \ldots & & \ldots \\ y_r x_1 & y_r x_2 & \ldots & y_r x_c \end{pmatrix} = \text{"matrix"}$$

Translation bra ↔ ket × real matrix **A**: $|u\rangle = \mathbf{A}|v\rangle \Leftrightarrow \langle u| = \langle v|\mathbf{A}^T$

$i$-th "**canonical basis**" bra vector: $\langle e_i| \equiv (0_1 0_2 \cdots 0_{i-1} \ 1_i \ 0_{i+1} \cdots 0_n)$

General "**orthonormal basis**" $\{B\} \equiv \{|b_i\rangle : i = 1..n\}$: $\langle b_i | b_j \rangle = \delta_{i,j}$

**Orthogonal** matrix **U**: $\mathbf{UU}^T = \mathbf{I}_n$, orthonormal bases $\{B\}, \{C\}$: $\mathbf{U} = \sum_i |b_i\rangle\langle c_i|$

Projector (matrix) onto the $i$-th canonical axis: $\mathbf{P}_i \equiv |e_i\rangle\langle e_i|$

Projector (matrix) onto any normalized ($\langle u|u\rangle = 1$) vector the $|u\rangle$: $\mathbf{P}_i \equiv |u\rangle\langle u|$

Component (vector) of $\langle X|$ along axis $\langle e_i|$: $\langle X|\mathbf{P}_i = \langle X|e_i\rangle \cdot \langle e_i| = \langle e_i| \ x_i$

"**Resolution of identity**" in any basis $\{B\}$: $\mathbf{I}_n = \sum_{i=1}^{n} |b_i\rangle\langle b_i|$

The above examples illustrate a rationale for Dirac notation: product expressions of the form with two "pointy" ends such as <...> are always scalars (numbers), while products of the form with two flat ends |...>...<...| are always matrices. Mixed ends products (those with one pointy and one flat end) such as <...| or |...> are always row or column vectors. Due to associativity of matrix products, these "object type rules" are valid however many other matrix or vector factors may be inside and outside of the selected sub-product of a given type. Also, the "resolution of identity" sums $\sum |b_i\rangle\langle b_i|$ can be freely inserted between any two adjacent bars ('flat ends') within a large product, further aiding in the breakup of longer chains of matrices into scalars. Such rules of thumb often suggest, purely visually, quick, mistake-proof simplifications e.g. any scalars spotted as …<...>… pattern can be immediately factored out.

## C. Hadamard Matrices and Walsh Functions

Hadamard matrix $\mathbf{H}_n$ (or $\mathbf{H}$) is a square $n \times n$ matrix defined by equation $\mathbf{H}_n \mathbf{H}_n^T = n\mathbf{I}_n$. Of interest here are the Sylvester type of $\mathbf{H}_n$ matrices characterized by the size constraint $n \equiv 2^d$. Under this constraint the $\mathbf{H}_n$ matrices can be constructed recursively (equivalent to Kronecker products of $\mathbf{H}_2$) as follows [14]:

$$\mathbf{H}_2 = \begin{pmatrix} 1 & 1 \\ 1 & -1 \end{pmatrix} \quad \mathbf{H}_{2n} = \begin{pmatrix} \mathbf{H}_n & \mathbf{H}_n \\ \mathbf{H}_n & -\mathbf{H}_n \end{pmatrix} \equiv \mathbf{H}_2 \otimes \mathbf{H}_n \equiv \mathbf{H}_2^{\otimes(d+1)} \tag{2.1}$$

The pattern of $\mathbf{H}_{32}$ ($d=5$) is shown in Fig. 2.2 with '-1' elements shown as '-' and coordinates in base 16.

```
           00 02 04 06 08 0A 0C 0E 10 12 14 16 18 1A 1C 1E
   0:00    1 1 1 1 1 1 1 1 1 1 1 1 1 1 1 1 1 1 1 1 1 1 1 1 1 1 1 1 1 1 1 1   00
   1:01    1 - 1 - 1 - 1 - 1 - 1 - 1 - 1 - 1 - 1 - 1 - 1 - 1 - 1 - 1 - 1 -   01
   2:02    1 1 - - 1 1 - - 1 1 - - 1 1 - - 1 1 - - 1 1 - - 1 1 - - 1 1 - -   02
   3:03    1 - - 1 1 - - 1 1 - - 1 1 - - 1 1 - - 1 1 - - 1 1 - - 1 1 - - 1   03
   4:04    1 1 1 1 - - - - 1 1 1 1 - - - - 1 1 1 1 - - - - 1 1 1 1 - - - -   04
   5:05    1 - 1 - - 1 - 1 1 - 1 - - 1 - 1 1 - 1 - - 1 - 1 1 - 1 - - 1 - 1   05
   6:06    1 1 - - - - 1 1 1 1 - - - - 1 1 1 1 - - - - 1 1 1 1 - - - - 1 1   06
   7:07    1 - - 1 - 1 1 - 1 - - 1 - 1 1 - 1 - - 1 - 1 1 - 1 - - 1 - 1 1 -   07
   8:08    1 1 1 1 1 1 1 1 - - - - - - - - 1 1 1 1 1 1 1 1 - - - - - - - -   08
   9:09    1 - 1 - 1 - 1 - - 1 - 1 - 1 - 1 1 - 1 - 1 - 1 - - 1 - 1 - 1 - 1   09
  10:0A    1 1 - - 1 1 - - - - 1 1 - - 1 1 1 1 - - 1 1 - - - - 1 1 - - 1 1   0A
  11:0B    1 - - 1 1 - - 1 - 1 1 - - 1 1 - 1 - - 1 1 - - 1 - 1 1 - - 1 1 -   0B
  12:0C    1 1 1 1 - - - - - - - - 1 1 1 1 1 1 1 1 - - - - - - - - 1 1 1 1   0C
  13:0D    1 - 1 - - 1 - 1 - 1 - 1 1 - 1 - 1 - 1 - - 1 - 1 - 1 - 1 1 - 1 -   0D
  14:0E    1 1 - - - - 1 1 - - 1 1 1 1 - - 1 1 - - - - 1 1 - - 1 1 1 1 - -   0E
  15:0F    1 - - 1 - 1 1 - - 1 1 - 1 - - 1 1 - - 1 - 1 1 - - 1 1 - 1 - - 1   0F
  16:10    1 1 1 1 1 1 1 1 1 1 1 1 1 1 1 1 - - - - - - - - - - - - - - - -   10
  17:11    1 - 1 - 1 - 1 - 1 - 1 - 1 - 1 - - 1 - 1 - 1 - 1 - 1 - 1 - 1 - 1   11
  18:12    1 1 - - 1 1 - - 1 1 - - 1 1 - - - - 1 1 - - 1 1 - - 1 1 - - 1 1   12
  19:13    1 - - 1 1 - - 1 1 - - 1 1 - - 1 - 1 1 - - 1 1 - - 1 1 - - 1 1 -   13
  20:14    1 1 1 1 - - - - 1 1 1 1 - - - - - - - - 1 1 1 1 - - - - 1 1 1 1   14
  21:15    1 - 1 - - 1 - 1 1 - 1 - - 1 - 1 - 1 - 1 1 - 1 - - 1 - 1 1 - 1 -   15
  22:16    1 1 - - - - 1 1 1 1 - - - - 1 1 - - 1 1 - - 1 1 - - 1 1 1 1 - -   16
  23:17    1 - - 1 - 1 1 - 1 - - 1 - 1 1 - - 1 1 - 1 - - 1 - 1 1 - 1 - - 1   17
  24:18    1 1 1 1 1 1 1 1 - - - - - - - - - - - - - - - - 1 1 1 1 1 1 1 1   18
  25:19    1 - 1 - 1 - 1 - - 1 - 1 - 1 - 1 - 1 - 1 - 1 - 1 1 - 1 - 1 - 1 -   19
  26:1A    1 1 - - 1 1 - - - - 1 1 - - 1 1 - - 1 1 - - 1 1 1 1 - - 1 1 - -   1A
  27:1B    1 - - 1 1 - - 1 - 1 1 - - 1 1 - - 1 1 - - 1 1 - 1 - - 1 1 - - 1   1B
  28:1C    1 1 1 1 - - - - - - - - 1 1 1 1 - - - - 1 1 1 1 1 1 1 1 - - - -   1C
  29:1D    1 - 1 - - 1 - 1 - 1 - 1 1 - 1 - - 1 - 1 1 - 1 - 1 - 1 - - 1 - 1   1D
  30:1E    1 1 - - - - 1 1 - - 1 1 1 1 - - - - 1 1 1 1 - - 1 1 - - - - 1 1   1E
  31:1F    1 - - 1 - 1 1 - - 1 1 - 1 - - 1 - 1 1 - 1 - - 1 1 - - 1 - 1 1 -   1F
           00 02 04 06 08 0A 0C 0E 10 12 14 16 18 1A 1C 1E
```

**Fig. 2.2**

From the construction eq. (2.1) of $\mathbf{H}_n$ (where $n \equiv 2^d$) it follows that $\mathbf{H}_n$ is a symmetric matrix:

$$\text{Symmetry:} \qquad H_{i,j} = H_{j,i} \tag{2.2}$$

Walsh function $\mathbf{U}_k(x)$ for $k=0..n$-1, $x=0..n$-1, is defined as the $k$-th row of $\mathbf{H}_n$. By virtue of $\mathbf{H}_n$ symmetry, eq. (2.2), the $k$-th column of $\mathbf{H}_n$ is also equal to $\mathbf{U}_k(x)$. The row and column forms of $\mathbf{U}_k(x)$ can also be used as the $n$-dimensional bra/ket or row/column vectors $\langle\mathbf{U}_k|$ and $|\mathbf{U}_k\rangle$. Some properties of $\mathbf{U}_k(x)$ are:

$$\text{Orthogonality:} \qquad \langle\mathbf{U}_j|\mathbf{U}_k\rangle = n \cdot \delta_{j,k} = \begin{cases} n & \text{for } j = k \\ 0 & \text{for } j \neq k \end{cases} \tag{2.3}$$

$$\text{Symmetry:} \qquad \mathbf{U}_k(x) = \mathbf{U}_x(k) \tag{2.4}$$

$$\text{Function values:} \qquad \mathbf{U}_k(x) = (-1)^{\sum_{\mu=0}^{d-1} k_\mu x_\mu} = (-1)^{\mathbb{P}(k\&x)} \tag{2.5}$$

$$\mathbf{U}_0(x) = (-1)^{\sum_{\mu=0}^{d-1} k_\mu x_\mu} = (-1)^{\sum_{\mu=0}^{d-1} 0 \cdot x_\mu} = (-1)^0 = 1, \quad \forall x \tag{2.6}$$

$$\sum_{x=0}^{n-1} \mathbf{U}_k(x) = 0 \quad \text{for } k = 1..n-1 \tag{2.7}$$

The exponent $\sum_{\mu=0}^{d-1} k_\mu x_\mu$ in eq. (2.5) uses binary digits $k_\mu$ and $x_\mu$ of $d$-bit integers $k$ and $x$. When this sum is even number $\mathbf{U}_k(x)$ is 1 and when the sum is odd number $\mathbf{U}_k(x)$ is -1. The second equality in eq. (2.5) expresses the same results via parity function $\mathbb{P}(k\&x)$, where $k\&x$ is a bitwise AND of integers $k$ and $x$. For example $\mathbf{U}_{14}(15)=(-1)$ from the table [Fig. 2.2](). Binary forms for $k$ and $x$ are: $k=14=01110$ and $x=15=01111$. The sum in the exponent is $\sum_{\mu=0}^{d-1} k_\mu x_\mu = 0\cdot 0+1\cdot 1+1\cdot 1+1\cdot 1+0\cdot 1 = 3 \Rightarrow \mathbf{U}_{14}(15) = (-1)^3 = (-1)^1 = -1$. The parity approach uses $k \& x = 01110 \& 01111 = 01110$ yielding exponent $\mathbb{P}(01110) = 0\wedge 1\wedge 1\wedge 1\wedge 0 = 1$ and $\mathbf{U}_{14}(15)=(-1)^1 = -1$ i.e. the same result as the one obtained via the sum formula.

For efficiency, the LH network computations use mostly binary (also called boolean) form of $\mathbf{U}_k$ and $\mathbf{H}_n$ denoted respectively as $\mathbf{W}_k$ and $[\mathbf{W}_n]$. When both forms are used in the same context, the $\mathbf{U}_k$ and $\mathbf{H}_n$ forms are referred to as algebraic forms. Binary form is obtained from the algebraic form via mappings $1 \to 0$ and $-1 \to 1$. Denoting algebraic values as $a$ and binary values as $b$, the translations between the two are:

| Algebraic | $a$ | -1 | 1 |
|---|---|---|---|
| Binary | $b$ | 1 | 0 |

$$b \equiv \frac{1-a}{2} \quad \text{and} \quad a = 1 - 2b \tag{2.8}$$

The symmetry eq. (2.4) and function values eq. (2.5) become for the binary form $\mathbf{W}_k(x)$:

$$\text{Symmetry:} \qquad \mathbf{W}_k(x) = \mathbf{W}_x(k) \tag{2.9}$$

$$\text{Function values:} \qquad \mathbf{W}_k(x) = \mathbb{P}\left(\sum_{\mu=0}^{d-1} k_\mu x_\mu\right) = \mathbb{P}(k\&x) \tag{2.10}$$

Binary Walsh functions $\mathbf{W}_k(x)$ are often treated as length $n$ bit strings, which for $k=1..n$-1 have exactly $n/2$ zeros and $n/2$ ones. In the bit string form one can perform bitwise Boolean operations on $\mathbf{W}_k$ as length $n$ bit strings. Their XOR property will be useful for the LH computations:

$$W_j \wedge W_k = W_{j \wedge k} \qquad (2.11)$$

i.e. the set $\{W_k\} \equiv \{W_k: k=0..n-1\}$ is closed with respect to bitwise XOR (denoted as ^) operation and it forms a group of *n*-bit strings isomorphic to the dyadic group $\mathbf{D}^d$ of their indices *k* (*d*-bit strings).

Figure 2.3 below shows the binary form of Hadamard (also called Walsh) matrix $[W_{32}]$ obtained via mapping eq. (2.8) from $H_{32}$ in Fig. 2.2 (binary 0's are shown as '-').

```
        00 02 04 06 08 0A 0C 0E 10 12 14 16 18 1A 1C 1E
 0:00   -  -  -  -  -  -  -  -  -  -  -  -  -  -  -  -  -  -  -  -  -  -  -  -  -  -  -  -  -  -  -  -   00
 1:01   -  1  -  1  -  1  -  1  -  1  -  1  -  1  -  1  -  1  -  1  -  1  -  1  -  1  -  1  -  1  -  1   01
 2:02   -  -  1  1  -  -  1  1  -  -  1  1  -  -  1  1  -  -  1  1  -  -  1  1  -  -  1  1  -  -  1  1   02
 3:03   -  1  1  -  -  1  1  -  -  1  1  -  -  1  1  -  -  1  1  -  -  1  1  -  -  1  1  -  -  1  1  -   03
 4:04   -  -  -  -  1  1  1  1  -  -  -  -  1  1  1  1  -  -  -  -  1  1  1  1  -  -  -  -  1  1  1  1   04
 5:05   -  1  -  1  1  -  1  -  -  1  -  1  1  -  1  -  -  1  -  1  1  -  1  -  -  1  -  1  1  -  1  -   05
 6:06   -  -  1  1  1  1  -  -  -  -  1  1  1  1  -  -  -  -  1  1  1  1  -  -  -  -  1  1  1  1  -  -   06
 7:07   -  1  1  -  1  -  -  1  -  1  1  -  1  -  -  1  -  1  1  -  1  -  -  1  -  1  1  -  1  -  -  1   07
 8:08   -  -  -  -  -  -  -  -  1  1  1  1  1  1  1  1  -  -  -  -  -  -  -  -  1  1  1  1  1  1  1  1   08
 9:09   -  1  -  1  -  1  -  1  1  -  1  -  1  -  1  -  -  1  -  1  -  1  -  1  1  -  1  -  1  -  1  -   09
10:0A   -  -  1  1  -  -  1  1  1  1  -  -  1  1  -  -  -  -  1  1  -  -  1  1  1  1  -  -  1  1  -  -   0A
11:0B   -  1  1  -  -  1  1  -  1  -  -  1  1  -  -  1  -  1  1  -  -  1  1  -  1  -  -  1  1  -  -  1   0B
12:0C   -  -  -  -  1  1  1  1  1  1  1  1  -  -  -  -  -  -  -  -  1  1  1  1  1  1  1  1  -  -  -  -   0C
13:0D   -  1  -  1  1  -  1  -  1  -  1  -  -  1  -  1  -  1  -  1  1  -  1  -  1  -  1  -  -  1  -  1   0D
14:0E   -  -  1  1  1  1  -  -  1  1  -  -  -  -  1  1  -  -  1  1  1  1  -  -  1  1  -  -  -  -  1  1   0E
15:0F   -  1  1  -  1  -  -  1  1  -  -  1  -  1  1  -  -  1  1  -  1  -  -  1  1  -  -  1  -  1  1  -   0F
16:10   -  -  -  -  -  -  -  -  -  -  -  -  -  -  -  -  1  1  1  1  1  1  1  1  1  1  1  1  1  1  1  1   10
17:11   -  1  -  1  -  1  -  1  -  1  -  1  -  1  -  1  1  -  1  -  1  -  1  -  1  -  1  -  1  -  1  -   11
18:12   -  -  1  1  -  -  1  1  -  -  1  1  -  -  1  1  1  1  -  -  1  1  -  -  1  1  -  -  1  1  -  -   12
19:13   -  1  1  -  -  1  1  -  -  1  1  -  -  1  1  -  1  -  -  1  1  -  -  1  1  -  -  1  1  -  -  1   13
20:14   -  -  -  -  1  1  1  1  -  -  -  -  1  1  1  1  1  1  1  1  -  -  -  -  1  1  1  1  -  -  -  -   14
21:15   -  1  -  1  1  -  1  -  -  1  -  1  1  -  1  -  1  -  1  -  -  1  -  1  1  -  1  -  -  1  -  1   15
22:16   -  -  1  1  1  1  -  -  -  -  1  1  1  1  -  -  1  1  -  -  -  -  1  1  1  1  -  -  -  -  1  1   16
23:17   -  1  1  -  1  -  -  1  -  1  1  -  1  -  -  1  1  -  -  1  -  1  1  -  1  -  -  1  -  1  1  -   17
24:18   -  -  -  -  -  -  -  -  1  1  1  1  1  1  1  1  1  1  1  1  1  1  1  1  -  -  -  -  -  -  -  -   18
25:19   -  1  -  1  -  1  -  1  1  -  1  -  1  -  1  -  1  -  1  -  1  -  1  -  -  1  -  1  -  1  -  1   19
26:1A   -  -  1  1  -  -  1  1  1  1  -  -  1  1  -  -  1  1  -  -  1  1  -  -  -  -  1  1  -  -  1  1   1A
27:1B   -  1  1  -  -  1  1  -  1  -  -  1  1  -  -  1  1  -  -  1  1  -  -  1  -  1  1  -  -  1  1  -   1B
28:1C   -  -  -  -  1  1  1  1  1  1  1  1  -  -  -  -  1  1  1  1  -  -  -  -  -  -  -  -  1  1  1  1   1C
29:1D   -  1  -  1  1  -  1  -  1  -  1  -  -  1  -  1  1  -  1  -  -  1  -  1  -  1  -  1  1  -  1  -   1D
30:1E   -  -  1  1  1  1  -  -  1  1  -  -  -  -  1  1  1  1  -  -  -  -  1  1  -  -  1  1  1  1  -  -   1E
31:1F   -  1  1  -  1  -  -  1  1  -  -  1  -  1  1  -  1  -  -  1  -  1  1  -  -  1  1  -  1  -  -  1   1F
        00 02 04 06 08 0A 0C 0E 10 12 14 16 18 1A 1C 1E
```

**Fig. 2.3**

## D. Error Correcting Codes

Error correcting coding (ECC) is a large variety of techniques for adding redundancy to messages in order to detect or correct errors in the decoding phase. Of interest for the LH network construction are the linear EC codes, which are the most developed and in practice the most important type of ECC [15],[16].

Message **X** is a sequence of $k$ symbols $x_1, x_2,\ldots, x_k$ from alphabet $\mathcal{A}$ of size $q \geq 2$ i.e. $x_i$ can be taken to be integers with values in interval $[0,q)$. EC code for **X** is a codeword **Y** which is a sequence $y_1, y_2,\ldots, y_n$ of $n > k$ symbols from $\mathcal{A}^*$. The encoding procedure translates all messages from some set $\{X\}$ of all possible messages into codewords from some set $\{Y\}$. For block codes the sizes of the sets $\{X\}$ and $\{Y\}$ are $q^k$ i.e. messages are arbitrary $k$-symbol sequences. The excess symbols $n-k > 0$ in **Y** represent coding redundancy or "check bits" that support detection or correction of errors during decoding of **Y** into **X**.

For ECC algorithmic purposes, the set $\mathcal{A}$ is augmented with additional mathematical structure, beyond merely that of a bare set of $q$ elements $\mathcal{A}$. The common augmentation is to consider symbols $x_i$ and $y_i$ to be elements of a Galois field $\mathbf{GF}(q)$ where $q \equiv p^m$ for some prime $p$ and some integer $m \geq 1$ (this condition on $q$ is a necessary condition in order to augment a bare set $\mathcal{A}$ into a finite field $\mathbf{F}_q$). Codewords **Y** are then a subset of all $n$-tuples $\mathbf{F}_q^n$ over the field $\mathbf{GF}(q)$. The $\mathbf{GF}(q)$ field arithmetic (i.e. the + and scalar ·) for the $n$-tuples $\mathbf{F}_q^n$ is done component-wise i.e. $\mathbf{F}_q^n$ is $n$-dimensional vector space $\mathbb{V}_n \equiv \mathbf{F}_q^n$ over $\mathbf{GF}(q)$.

Linear EC codes are a special case of the above $n$-tuple $\mathbf{F}_q^n$ structure of codewords, in which the set $\{Y\}$ of all codewords is a $k$-dimensional vector *subspace* (or **span**) $\mathbb{S}(k,n,q)$ of $\mathbb{V}_n$. Hence, if two $n$-tuples $\mathbf{Y}_1$ and $\mathbf{Y}_2$ are codewords, then the $n$-tuple $\mathbf{Y}_3 = \mathbf{Y}_1 + \mathbf{Y}_2$ is also a codeword. The number of distinct codewords **Y** in $\mathbb{S}(k,n,q)$ is $|\mathbb{S}(k,n,q)| = q^k$. This linear code is denoted in ECC convention as $[n,k]_q$ code, or just $[n,k]$ code when $q$ is understood from the context or otherwise unimportant in a context.

A particular $[n,k]$ code can be defined by specifying $k$ linearly independent $n$-dimensional row vectors $\langle g_i | = (g_{i,1}\; g_{i,2}\; \ldots\; g_{i,n})$ for $i=1..k$, which are used to define the $k \times n$ "**generator matrix**" $[G]$ of the $[n,k]$ code as follows ([16] p. 84):

$$[\mathbf{G}] \equiv \sum_{i=1}^{k} |e_i\rangle\langle g_i| = \begin{pmatrix} \langle g_1 | \\ \ldots \\ \langle g_k | \end{pmatrix} = \begin{pmatrix} g_{1,1} & g_{1,2} & \ldots & g_{1,n} \\ \ldots & \ldots & \ldots & \ldots \\ g_{k,1} & g_{k,2} & \ldots & g_{k,n} \end{pmatrix} \quad (2.20)$$

Encoding of a message $\mathbf{X} \equiv \langle \mathbf{X}| \equiv (x_1, x_2, \ldots, x_k)$ into the codeword $\mathbf{Y} \equiv \langle \mathbf{Y}| \equiv (y_1, y_2, \ldots, y_n)$ is:

$$\langle \mathbf{Y}| \equiv \langle \mathbf{X}|[\mathbf{G}] = \left(\sum_{j=1}^{k} x_j \langle e_j|\right)\left(\sum_{i=1}^{k} |e_i\rangle\langle g_i|\right) = \sum_{i,j=1}^{k} x_j \delta_{j,i} \langle g_i| = \sum_{i=1}^{k} x_i \langle g_i| \quad (2.21)$$

Individual component (symbol) $y_s$ (where $s=1..n$) of the codeword **Y** is then via eqs. (2.20)-(2.21):

---

[*] More generally message **X** and codeword **Y** can use different alphabets, but this generality merely complicates exposition without adding anything useful for the LH construction.

$$y_s \equiv \langle Y|e_s\rangle = \sum_{i=1}^{k} x_i \langle g_i|e_s\rangle = \sum_{i=1}^{k} x_i g_{i,s} \tag{2.22}$$

The $k\times n$ matrix $[G_{k,n}]$ is called **systematic generator** iff the original message $X = x_1, x_2,\ldots, x_k$ occurs as a substring of the output codeword $Y$. The systematic generators $[G]$ combine a $k\times k$ identity matrix $I_k$ as a sub-matrix of $[G]$ i.e. $[G]$ typically has a form $[I_k \mid A_{k,n-k}]$ or $[A_{k,n-k} \mid I_k]$, yielding unmodified substring $X$ as a prefix or a suffix of $Y$, which simplifies encoding and decoding operations. The remaining $n$-$k$ symbols of $Y$ are then called parity check symbols.

The choice of vectors $\langle g_i|$ used to construct $[G]$ depends on type of errors that the $[n,k]$ code is supposed to detect or correct. For the most common assumption in ECC theory, the independent random errors for symbols of codeword $Y$, the best choice of $\langle g_i|$ are those that maximize the minimum Hamming distance $\Delta(Y_1,Y_2)$ among all pairs $(Y_1,Y_2)$ of codewords. Defining minimum codeword distance via:

$$\Delta \equiv \min\{\Delta(Y_1, Y_2) \mid \forall\, Y_1, Y_2 \in \mathbb{S}(k,n,q) \text{ and } Y_1 \neq Y_2\} \tag{2.24}$$

the $[n,k]_q$ code is often denoted as $[n,k,\Delta]_q$ or $[n,k,\Delta]$ code. The optimum choice for vectors $\langle g_i|$ maximizes $\Delta$ for given $n$, $k$ and $q$. The tables of optimum and near optimum $[n,k,\Delta]_q$ codes have been computed over decades for wide ranges of free parameters $n$, $k$ and $q$ (e.g. see web repository [17]).

Fig. 2.4 ([16] p. 34) illustrates optimum $[7,4,3]_2$ code i.e. a systematic binary code with $n=7$ bit codewords each containing 3 parity check bits, $k=4$ message bits (appearing as suffix in the codeword $Y$), with minimum distance $\Delta=3$, thus capable of correcting all 1-bit errors and detecting all 2-bit errors.

$$[G_{4,7}] = \begin{pmatrix} 1 & 1 & 0 & \mathbf{1} & 0 & 0 & 0 \\ 0 & 1 & 1 & 0 & \mathbf{1} & 0 & 0 \\ 1 & 1 & 1 & 0 & 0 & \mathbf{1} & 0 \\ 1 & 0 & 1 & 0 & 0 & 0 & \mathbf{1} \end{pmatrix}$$

**Fig. 2.4**

Quantity closely related to $\Delta$, and of importance for LH construction, is the minimum non-zero codeword weight $w_{min}$ defined via Hamming weight $\langle Y\rangle$ (the number of non-zero symbols in $Y$) as follows:

$$\mathbf{w}_{min} \equiv \min\{\langle Y\rangle : \big(Y \in \mathbb{S}(k,n,q)\big) \text{ and } (Y \neq 0)\} \tag{2.25}$$

The property of $w_{min}$ (cf. Theorem 3.1, p. 83 in [16]) of interest is that for any linear code $[n,k,\Delta]_q$:

$$\mathbf{w}_{min} = \Delta \tag{2.26}$$

Hence, the ***construction of optimal*** $[n,k,\Delta]_q$ *codes* (maximizing $\Delta$) is a problem of finding $k$-dimensional subspace $\mathbb{S}(k,n,q)$ of an $n$-dimensional space $F_q^n$ which maximizes $w_{min}$. Note also that since any set of $k$ linearly independent vectors $\langle g_i|$ (a basis) from $\mathbb{S}(k,n,q)$ generates (spans) the same space $\mathbb{S}(k,n,q)$ of $q^k$ vectors $Y$, $w_{min}$ and $\Delta$ are ***independent of the choice of the basis*** $\{\langle g_i|: i=1..k\}$. Namely by virtue of uniqueness of expansion of all $q^k$ vectors $Y \in \mathbb{S}(k,n,q)$ in any basis and pigeonhole principle, the change of basis merely permutes the mapping $X\to Y$, retaining exactly the same set of $q^k$ vectors of $\mathbb{S}(k,n,q)$.

## E. Graphs: Terms and Notation

- $\Gamma(V,E)$ — Graph $\Gamma$ with vertices $V=\{v_1, v_2, \ldots v_n\}$ and edges $E=\{\varepsilon_1, \varepsilon_2, \ldots \varepsilon_c\}$
- degree of $v$ — Number of edges (links) connected to node $v$
- $\Gamma_1 \square \Gamma_2$ — Cartesian product of graphs $\Gamma_1$ and $\Gamma_2$ (syn. "product graph")
- $\Gamma^{\square n}$ — (Cartesian) $n$-th power of graph $\Gamma$
- $\varepsilon_k = (v_i, v_j)$ — Edge $\varepsilon_k$ connects vertices $v_i$ and $v_j$
- $v_i \sim v_j$ — Vertices $v_i$ and $v_j$ are connected
- $v_i \not\sim v_j$ — Vertices $v_i$ and $v_j$ are not connected
- $[\mathbf{A}]$ — Adjacency matrix of a graph: $[\mathbf{A}]_{i,j} \equiv A(i,j) \equiv [v_i \sim v_j]$: 1 if $v_i \sim v_j$, 0 if $v_i \not\sim v_j$. Number of ones on a row $r$ (or column $c$) is the degree of node $r$ (or $c$)
- $A(i,j) = A(j,i)$ — Symmetry property of $[\mathbf{A}]$ (for undirected graphs)
- $\mathbb{C}_n$ — Cycle graph: A ring with $n$ vertices (syn. $n$-ring)
- $\mathbb{P}_n$ — Path graph: $n$-ring with one link broken i.e. a line with $n$ vertices (syn. $n$-path)
- $\mathbb{Q}_d$ — $d$-dimensional hypercube (syn. $d$-cube): $(\mathbb{P}_2)^{\square d} = \mathbb{P}_2 \square \mathbb{P}_2 \square \ldots \square \mathbb{P}_2$ ($d$ times)
- $\mathbb{FQ}_d$ — Folded $d$-cube: $d$-cube with extra link on each long diagonal (see Fig. 4.4)

**Cayley Graph** $Cay(\mathbf{G}_n, \mathbf{S}_m)$, where: $\mathbf{G}_n$ is a group with $n$ elements $\{g_1 \equiv \mathbf{I}_0, g_2, \ldots g_n\}$ and $\mathbf{S}_m$, called **generator set**, is a subset of $\mathbf{G}_n$ with $m$ elements: $\mathbf{S}_m = \{h_1, h_2, \ldots h_m\}$ such that (cf. [18] chap. 5):

(i) for any $h \in \mathbf{S}_m \Rightarrow h^{-1} \in \mathbf{S}_m$ (i.e. $\mathbf{S}_m$ contains inverse of any of its elements)
(ii) $\mathbf{S}_m$ does *not* contain identity element (denoted as $\mathbf{I}_0$) $g_1$ of $\mathbf{G}_n$ [*]

**Construction:** Vertex set $V$ of $Cay(\mathbf{G}_n, \mathbf{S}_m)$ is $V \equiv \{g_1, g_2, \ldots g_n\}$ and the edge set is $E \equiv \{(g_i, g_i \cdot h_s), \forall i, s\}$. In words, each vertex $g_i$ is connected to $m$ vertices $g_i \cdot h_s$ for $s = 1..m$. Generating elements $h_s$ are called here "hops" since for identity element $g_1 \equiv \mathbf{I}_0$ ("root node") their group action is precisely the single hop transition from the root node $g_1$ to its 1-hop neighbors $h_1, h_2, \ldots h_m \in V(\mathbf{G}_n)$.

The construction of $\mathbb{Q}_3 = Cay(\mathbf{D}^3, \mathbf{S}_3)$ is illustrated in Fig. 2.5. Group is the 8 element Dyadic group $\mathbf{D}^3$ and the 3 generators $h_1 = 001$, $h_2 = 010$ and $h_3 = 100$ are shown with arrows indicating the group action (XORs node labels with generators; all labels are in binary) on vertex $v_1 = 000$. The resulting graph is a 3-cube.

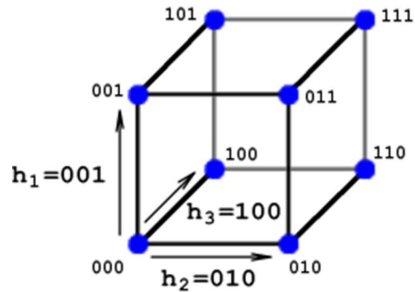

**Fig. 2.5**

---

[*] The requirement for inverse $h^{-1}$ to be in $\mathbf{S}_m$ applies to *undirected* Cayley graphs, not to directed graphs. The exclusion of identity $\mathbf{S}_m$ applies to graphs that have no self-loops of a node to itself (i.e. a vertex $v \sim v$). These restrictions are not essential but mere conveniences of the concrete implementation.

## F. Properties of Matrices

This section lists several results about matrices (cf. [19]) needed in LH construction. All matrices below will be assumed to be real (rather than complex valued matrices).

**M$_1$)** Square $n \times n$ real matrix **A** is called normal matrix ([19] p. 100) iff it satisfies relation:

$$\mathbf{A}\mathbf{A}^T = \mathbf{A}^T\mathbf{A} \tag{2.40}$$

This implies that any *symmetrical* (real) matrix **S** is *normal* matrix (since $\mathbf{S}=\mathbf{S}^T$, hence $\mathbf{S}\mathbf{S}^T=\mathbf{S}^2=\mathbf{S}^T\mathbf{S}$).

**M$_2$)** Any real, symmetrical $n \times n$ matrix [**S**] has $n$ real eigenvalues $\lambda_i$ ($i=1..n$) and the $n$ corresponding orthonormal eigenvectors: $|v_i\rangle$ for $i=1..n$ (cf. [19] p.101):

$$[\mathbf{S}]|v_i\rangle = \lambda_i |v_i\rangle \quad for \ i = 1..n \tag{2.41}$$

$$\langle v_i | v_j \rangle = \delta_{i,j} \tag{2.42}$$

**M$_3$)** Since set $\{|v_i\rangle\}$ is a complete orthonormal set of vectors (a basis in $\mathbb{V}_n$), any [**S**] from (M$_2$) can be diagonalized via an orthogonal $n \times n$ matrix" [**U**] (orthogonal matrix is defined via condition [**U**][**U**$^T$]=**I**$_n$) which can be constructed as follows (applying eqs. (2.41)-(2.42)):

$$[\mathbf{U}] \equiv \sum_{i=1}^{n} |e_i\rangle \langle v_i| \tag{2.43}$$

$$[\mathbf{U}][\mathbf{S}][\mathbf{U}^T] = \left(\sum_{i=1}^{n} |e_i\rangle \langle v_i|\right)[\mathbf{S}]\left(\sum_{j=1}^{n} |v_j\rangle \langle e_j|\right) = \sum_{i,j=1}^{n} |e_i\rangle \langle e_j| \lambda_j \delta_{i,j} = \sum_{i=1}^{n} \lambda_i |e_i\rangle \langle e_i| \tag{2.44}$$

The final sum in (2.44) is a diagonalized form of [**S**], with $\lambda_i$'s along main diagonal and 0's elsewhere.

**M$_4$)** A set of $m$ symmetric, pairwise commuting matrices $\mathcal{F}_m \equiv \{\mathbf{S}_r: \mathbf{S}_r\mathbf{S}_t = \mathbf{S}_t\mathbf{S}_r \text{ for } t,r=1..n\}$ is called **commuting family** (cf. [19] p. 51). For each commuting family $\mathcal{F}_m$ there is an orthonormal set of $n$ vectors (eigenbasis in $\mathbb{V}_n$) $\{|v_i\rangle\}$ which are simultaneously eigenvectors of all $\mathbf{S}_r \in \mathcal{F}_m$ (cf. [19] p. 52).

**M$_5$)** Labeling the $n$ eigenvalues of the symmetric matrix **S** from (M$_1$) as: $\lambda_{\min} \equiv \lambda_1 \leq \lambda_2 \leq \cdots \leq \lambda_n \equiv \lambda_{\max}$, then the following equalities hold (Rayleigh-Ritz theorem, [19] p. 176):

$$\lambda_{min} \equiv \lambda_1 = \min\left\{\frac{\langle X|\mathbf{S}|X\rangle}{\langle X|X\rangle}, for \ |X\rangle \in \mathbb{V}_n \ and \ |X\rangle \neq 0\right\} \tag{2.45}$$

$$\lambda_{max} \equiv \lambda_n = \max\left\{\frac{\langle X|\mathbf{S}|X\rangle}{\langle X|X\rangle}, for \ |X\rangle \in \mathbb{V}_n \ and \ |X\rangle \neq 0\right\} \tag{2.46}$$

# 3. Network Optimization Problems

Networks considered here consist of *n* "switches" (or nodes) of radix (number of ports per switch) $R_i$ for the *i*-th switch, where $i = 1..n$. The network thus has the total of $P_T = \sum_i R_i$ ports. Some number of ports $P_I$ is used for internal connections between switches ("*topological ports*") leaving $P = P_T - P_I$ ports free ("*external ports*"), available for use by servers, routers, storage,… etc. The number of cables $C_I$ used by the internal connections is $C_I = P_I/2$. For regular networks (graphs), those in which all nodes have the same number of topological links per node *m* (i.e. *m* is a node degree), it follows $P_I = n \cdot m$.

The network capacity or throughput is commonly characterized via the bisection (bandwidth) which is defined in the following manner: network is partitioned into two equal subsets (**equipartition**) $S_1 + S_2$ so that each subset contains *n*/2 nodes (within $\pm 1$ for odd *n*). The total **number of links** connecting $S_1$ and $S_2$ is called a **cut** for partition $S_1+S_2$. **Bisection B** is defined as the smallest cut (**min-cut**) for all possible equipartitions $S_1+S_2$ of the network. Fig. 3.1 illustrates this definition on an 8 node network with **B**=2.

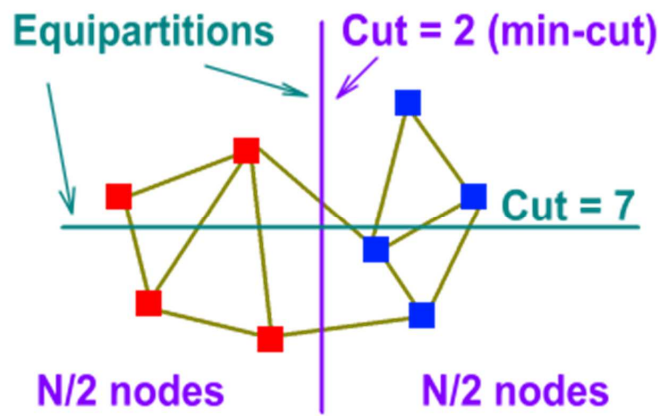

**Fig. 3.1**

Bisection is thus an absolute measure of the network bottleneck throughput. A related commonly used relative throughput measure is the **network oversubscription** ϕ defined by considering the **P**/2 free ports in each min-cut half, $S_1$ and $S_2$, with each port sending and receiving at its maximum capacity to/from the ports in the opposite half. The maximum traffic that can be sent in each direction this way without overloading the network is **B** link (port) capacities since that's how many links the bisection has between the halves. Any additional demand that free ports are capable of generating is thus considered to be an "oversubscription" of the network. Hence, the oversubscription ϕ is defined as the ratio:

$$\phi \equiv \frac{P/2}{B} \quad (3.1)$$

The performance comparisons between network topologies, such as [1]-[5], [9]-[10], typically use non-oversubscribed networks (ϕ=1) and compare the costs in terms of number of switches *n* of common radix **R** and number of internal cables $C_I$ used in order to obtain a given target number of free ports **P**. Via eq. (3.1), that is equivalent to comparing the costs **n** and $C_I$ needed to obtain a common target bisection **B**.

Therefore, the fundamental underlying problem is how to maximize **B** given the number of switches *n* each using some number of topological ports per switch *m* (node degree). This in turn breaks down into two sub-problems:

   **(i)** Compute bisection **B** for given network
   **(ii)** Modify/select links which maximize **B** computed via (i)

For general networks (graphs), both sub-problems are computationally intractable i.e. NP-complete problems. For example, the 'easier' of the two tasks[*], (i), finding the graph equipartition $H_0+H_1$ which has the minimum number of links between the two halves, in general case would have to examine every possible equipartition $H_0+H_1$ and in each case count the links between the two, then pick the one with the lowest count. Since there are $C(n, n/2) \simeq 2^n/\sqrt[2]{\pi n/2}$ ways to split the set of *n* nodes into two equal halves, the exact brute force solution has exponential complexity. The problem with approximate bisection algorithms is the poor solution quality as network size increases – the polynomial complexity algorithms bisection applied to general graphs cannot guarantee to find an approximate cut even to within merely a constant factor from the actual minimum cut as *n* increases. And without an accurate enough measure of network throughput, the subtask (ii) cannot even begin to optimize the links.

Additional problem with (ii) becomes apparent even for small networks, such as those with few dozen nodes, for which one can compute exact **B** via brute force and also compute the optimum solution by examining all combinations of the links. Namely, a greedy approach for solving (ii), successively computes **B** for all possible addition of the next link, then picks the link which produces the largest increment of **B** among all possible additions. That procedure continues until the target number of links per node is reached. The numerical experiments on small networks show that in order to get the optimum network in step *m* → *m*+1 links per node, one often needs to replace one or more existent links as well, the links which were required for optimum at previous smaller values of *m*.

In addition to bandwidth optimization for a given number of switches and cables, the latency, average or maximum (diameter), is another property that is often a target of optimization. Unlike the **B** optimization, where an optimum solution dramatically reduces network costs, yielding ~2-5 fewer switches and cables compared to conventional and approximate solutions, the latency is far less sensitive to the distinction between the optimal and approximate solutions, with typical advantage factors of only 1.2-1.5. Hence, the primary optimization objective of LH networks is the bisection, while latency is a secondary objective.

---

[*] Since (ii) requires multiple evaluations of (i) as the algorithm (ii) iterates/searches for the optimum **B**.

# 4. Construction of Long Hop Networks

The LH networks are direct networks constructed using general Cayley graphs *Cay*(**G**$_n$, **S**$_m$) for the topology of the switching network. The implemented variant of LH networks belongs to the most general hypercube-like networks, with uniform number of external (**E**) and topological (*m*) ports per switch (where **E**+*m*=**R**='switch radix'), which retain the vertex and edge symmetries of the regular *d*-cube $\mathbb{Q}_d$. The resulting LH network in that case is a Cayley graph of type *Cay*($\mathbf{Z}_2^d$, **S**$_m$) with *m* > *d*+1 (this choice of *m* excludes *d*-cube $\mathbb{Q}_d$ which has *m* = *d* and folded *d*-cube $\mathbb{F}\mathbb{Q}_d$ with *m* = *d*+1). It will become evident that the construction method shown on $\mathbf{Z}_2^d$ example applies directly to the general group $\mathbf{Z}_q^d$ with *q* > 2. For *q* > 2, the resulting *Cay*($\mathbf{Z}_q^d$, **S**$_m$) is the most general LH type construction of a *d*-dimensional hyper-torus-like or flattened butterfly-like network of extent *q* (which is equivalent to a hyper-mesh-like network with cyclic boundary conditions). The implementation will use *q* = 2, since $\mathbf{Z}_2^d$ is the most optimal choice from practical perspective due to the shortest latency (average and max), highest symmetry, simplest forwarding and routing, simplest job partitioning (e.g. for multi-processor clusters), easiest and most economical wiring in the $\mathbf{Z}_q^d$ class.

Following the overall task breakdown in section 3, the LH construction proceeds in two main phases:

(i) Constructing a method for efficient computation of the exact bisection **B**
(ii) Computing the optimal set of *m* links (hops) **S**$_m$ per node maximizing this **B**

For the sake of clarity, the main phases are split further into smaller subtasks, each described in the sections that follow.

## A. Generators and Adjacency Matrix

Network built on *Cay*($\mathbf{Z}_q^d$, **S**$_m$) graph has *n* = *q*$^d$ vertices (syn. nodes), hence for *q* = 2 used in the practical LH implementation *n* = **2**$^d$ nodes. These *n* nodes make the *n* element vertex set **V**={*v*$_0$,*v*$_2$,... *v*$_{n-1}$}[*].

### 1) Node labels and group operation table

The nodes *v*$_i$ are labeled using *d*-tuples in alphabet of size *q*: *v*$_i$ ≡ *i* ∈ {0,1,... *n*-1} expressed as *d*-digit integers in base *q*. The group operation, denoted as ⊕, is *not* the same as integer addition mod *n* but rather it is the component-wise addition modulo *q* done on *d* components separately. For *q* = 2, this is equivalent to a bitwise XOR operation between the *d*-tuples, as illustrated in Fig. 2.1 which shows the full $\mathbf{Z}_2^d$ group operation table for *d* = 4.

---

[*] We are using 0-based subscripts since we need to do modular arithmetic with them.

Fig. 4.1 illustrates analogous $\mathbf{Z}_3^d$ group operation table for *d*=2 (*q*=3) hence there are *n*=3²=9 group elements and the operations table has *n*×*n* = 9×9 = 81 entries. The 2-digit labels have digits which are from {0,1,2}. The *n* rows and *n* columns are labeled using 2-digit node labels. Table entry at row *r* and column *c* contains result of *r*⊕*c* (component-wise addition mod *q*=3). For example, the 3rd row labeled 02, and the 6-th column labeled 12, yield table entry 02⊕12 = (0+1)%3, (2+2)%3 =1,1 = 11.

```
        00 01 02 10 11 12 20 21 22
00:     00 01 02 10 11 12 20 21 22    :00
01:     01 02 00 11 12 10 21 22 20    :01
02:     02 00 01 12 10 11 22 20 21    :02
10:     10 11 12 20 21 22 00 01 02    :10
11:     11 12 10 21 22 20 01 02 00    :11
12:     12 10 11 22 20 21 02 00 01    :12
20:     20 21 22 00 01 02 10 11 12    :20
21:     21 22 20 01 02 00 11 12 10    :21
22:     22 20 21 02 00 01 12 10 11    :22
        00 01 02 10 11 12 20 21 22
```

**Fig. 4.1**

It can be noted in Fig. 4.1 for $\mathbf{Z}_3^d$ and in [Fig. 2.1](#) for $\mathbf{Z}_2^d$ that each row *r* and column *c* contains all *n* group elements, but in a unique order. The 0-th row or 0-th column contain the unmodified *r* and *c* values since the 'identity element' $\mathbf{I}_0$=0. Both tables are symmetrical since the operation *r*⊕*c* = *c*⊕*r* is symmetrical (which is a characteristic of the abelian group $\mathbf{Z}_q^d$).

## 2) Construction of adjacency matrix [A]

Generator set $\mathbf{S}_m$ contains *m* "hops" $h_1, h_2,\ldots h_m$ (they are also elements of the group $\mathbf{G}_n$ in $Cay(\mathbf{G}_n, \mathbf{S}_m)$), which can be viewed as the labels of the *m* nodes to which the "root" node, $v_0 \equiv 0$ is connected. Hence, the row *r*=0 of the adjacency matrix [A] has *m* ones, at columns **A**(0,*h*) for *m* hops *h* ∈ $\mathbf{S}_m$ and 0 elsewhere. Similarly, the column *c*=0 has *m* ones at rows **A**(*h*,0) for *m* hops *h* ∈ $\mathbf{S}_m$ and 0 elsewhere. In a general case, some row *r*=*y* has *m* ones at columns **A**(*y*,*y*⊕*h*) for *h* ∈ $\mathbf{S}_m$ and 0 elsewhere. Similarly a column *c*=*x* has *m* ones at rows **A**(*x*⊕*h*,*x*) for *h* ∈ $\mathbf{S}_m$ and 0 elsewhere. Denoting contributions of a single generator *h* ∈ $\mathbf{S}_m$ to the adjacency matrix [A] as a matrix **T**(*h*), these conclusions can be written more compactly via [Iverson brackets](#) and bitwise OR operator '|' as:

$$\mathbf{T}(a)_{i,j} \equiv [i \oplus a = j] \mid [j \oplus a = i] \quad a \in \mathbf{G}_n \tag{4.1}$$

$$[\mathbf{A}] = \sum_{h \in \mathbf{S}_m} \mathbf{T}(h) = \sum_{s=1}^{m} \mathbf{T}(h_s) \tag{4.2}$$

Note that eq. (4.1) defines **T**(*a*) for any element *a* (or vertex) of the group $\mathbf{G}_n$. Since the right hand side expression in eq. (4.1) is symmetric in *i* and *j* it follows that **T**(*a*) is a symmetric matrix, hence it has real, complete eigenbasis:

$$\mathbf{T}(a)_{i,j} = \mathbf{T}(a)_{j,i} \tag{4.3}$$

For the group $G_n = Z_2^d$, the group operator ⊕ becomes regular XOR '^', simplifying eq. (4.1) to:

$$T(a)_{i,j} \equiv [i \wedge j = a], \quad a \in Z_2^d \qquad (4.4)$$

Fig. 4.2 illustrates the **T**(*a*) matrices for *q*=2, *d*=3, *n*=8 and all group elements 0..7. For given *a*=0..7, value 1 is placed on row *r* and column *c* iff *r^c = a*, and 0 otherwise (0s are shown as '-').

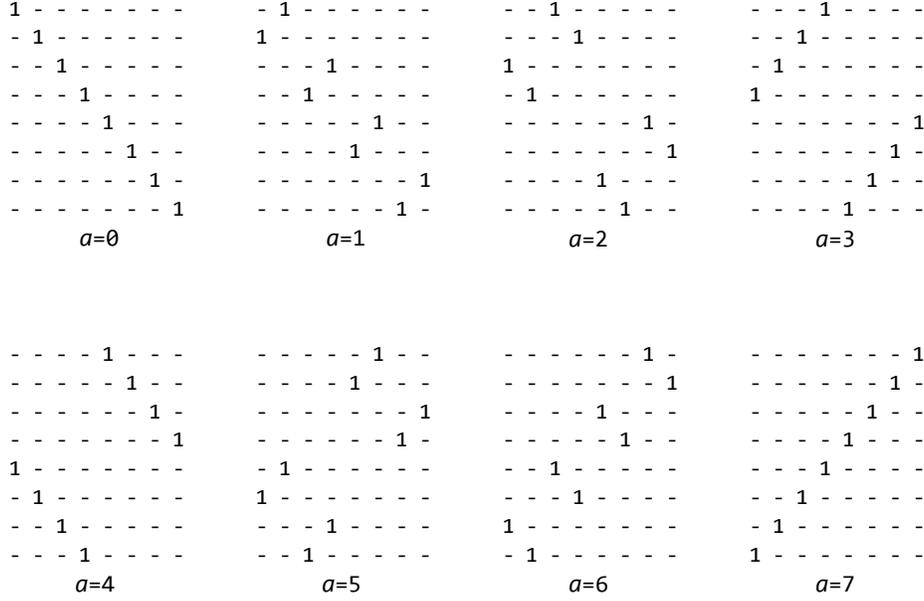

**Fig. 4.2**

Fig. 4.3-(a) shows the 8×8 adjacency matrix [**A**] obtained for the generator set $S_4 \equiv \{1, 2, 4, 7\}_{hex} \equiv \{001, 010, 100, 111\}_{bin}$ by adding the 4 generators from Fig. 4.2: [**A**] = **T**(1)+**T**(2)+**T**(4)+**T**(7), via eq. (4.2). For pattern clarity, values 0 are shown as '-'. Fig. 4.3-(b) shows the indices of the 4 generators (1,2,3,4) which contributed 1 to a given element of [**A**] in Fig. 4.3-(a).

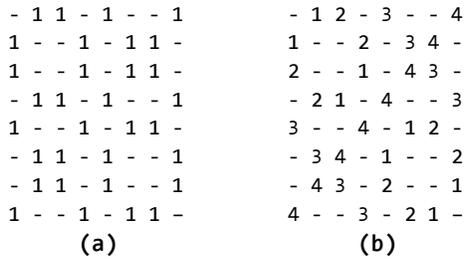

**Fig. 4.3**

Fig. 4.4 shows the resulting 8-node network (folded 3-cube, $\mathbb{FQ}_3$). Actions (bitwise XOR) of the 4 generators $\mathbf{T}(a) \in \{001, 010, 100, 111\}_{bin}$ on the node 000 are indicated by the arrows pointing to the target vertex. All other links are shown without arrows. The total number of links is $\mathbf{C} = \mathbf{n} \cdot \mathbf{m}/2 = 8 \cdot 4/2 = 16$, which can be observed directly in the figure.

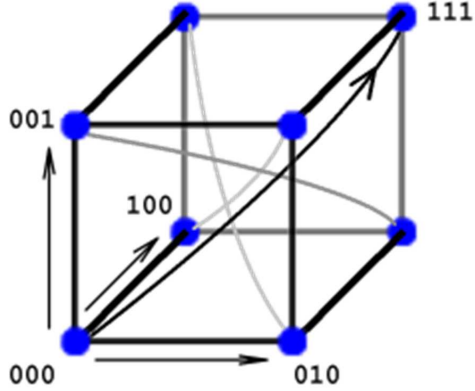

**Fig. 4.4**

### 3) Eigenvectors of T(*a*) and [A]

To solve the eigen-problem of [A], couple additional properties of $\mathbf{T}(a)$ are derived from eq. (4.4) (using x^x=0 and x^y=y^x):

$$\left(\mathbf{T}(a)\mathbf{T}(b)\right)_{i,j} = \sum_{k=0}^{n-1} \mathbf{T}(a)_{i,k} \mathbf{T}(b)_{k,j} = \sum_{k=0}^{n-1} [k = i\wedge a][k = j\wedge b] =$$

$$= [i\wedge a = j\wedge b] = [i\wedge j = a\wedge b] = \mathbf{T}(a\wedge b)_{i,j} \quad \Longrightarrow$$

$$\mathbf{T}(a)\mathbf{T}(b) = \mathbf{T}(a\wedge b) \tag{4.5}$$

$$\mathbf{T}(a)\mathbf{T}(b) = \mathbf{T}(a\wedge b) = \mathbf{T}(b\wedge a) = \mathbf{T}(b)\mathbf{T}(a) \tag{4.6}$$

Eq. (4.5) shows that $\mathbf{T}(a)$ matrices are a representation of the group $\mathbf{G}_n$ and eq. (4.6) that they commute with each other. Since via eq. (4.2), [A] is the sum of $\mathbf{T}(a)$ matrices, then [A] commutes with all $\mathbf{T}(a)$ matrices as well. Therefore, since they are all also symmetric matrices, the entire set $\{$ [A], $\mathbf{T}(a)$ $\forall a\}$ has a common eigenbasis (via result (M_4) in section 2.F). The next sequence of equations shows that Walsh functions viewed as *n*-dimensional vectors $|\mathbf{U}_k\rangle$ are the eigenvectors for $\mathbf{T}(a)$ matrices. Using eq. (4.4) for the matrix elements of the $\mathbf{T}(a)$, the action of $\mathbf{T}(a)$ on Walsh ket vector $|\mathbf{U}_k\rangle$ yields for the $i$-th component of the resulting vector:

$$(\mathbf{T}(a)|\mathbf{U}_k\rangle)_i = \sum_{j=0}^{n-1} \mathbf{T}(a)_{i,j} \mathbf{U}_k(j) = \sum_{j=0}^{n-1} [j = i \wedge a] \mathbf{U}_k(j) = \mathbf{U}_k(i \wedge a) \tag{4.7}$$

The result $\mathbf{U}_k(i \wedge a)$ is transformed via eq. (2.5) for the general function values of $\mathbf{U}_k(x)$:

$$\mathbf{U}_k(i \wedge a) = (-1)^{\sum_{\mu=0}^{d-1} k_\mu (i \wedge a)_\mu} = (-1)^{\sum_{\mu=0}^{d-1} k_\mu a_\mu + \sum_{\mu=0}^{d-1} k_\mu i_\mu} =$$

$$= (-1)^{\sum_{\mu=0}^{d-1} k_\mu a_\mu} \cdot (-1)^{\sum_{\mu=0}^{d-1} k_\mu i_\mu} = \mathbf{U}_k(a)\mathbf{U}_k(i) = \mathbf{U}_k(a)(|\mathbf{U}_k\rangle)_i \tag{4.8}$$

Collecting all $n$ components of the left side of eq. (4.7) and right side of eq. (4.8) yields in vector form:

$$\mathbf{T}(a)|\mathbf{U}_k\rangle = \mathbf{U}_k(a)|\mathbf{U}_k\rangle \tag{4.9}$$

Hence, the orthogonal basis set $\{|\mathbf{U}_k\rangle, k=0..n-1\}$ is the common eigenbasis for all $\mathbf{T}(a)$ matrices and for the adjacency matrix $[\mathbf{A}]$. The $n$ eigenvalues for $\mathbf{T}(a)$ are Walsh function values $\mathbf{U}_k(a)$, $k=0..n-1$. The eigenvalues for $[\mathbf{A}]$ are obtained by applying eq.(4.9) to the expansion of $[\mathbf{A}]$ via $\mathbf{T}(h)$, eq. (4.2):

$$[\mathbf{A}]|\mathbf{U}_k\rangle = \sum_{s=1}^{m} \mathbf{T}(h_s) |\mathbf{U}_k\rangle = \left(\sum_{s=1}^{m} \mathbf{U}_k(h_s)\right) \cdot |\mathbf{U}_k\rangle \equiv \lambda_k |\mathbf{U}_k\rangle \tag{4.10}$$

$$\text{where:} \quad \lambda_k \equiv \sum_{s=1}^{m} \mathbf{U}_k(h_s) \tag{4.11}$$

Since $\mathbf{U}_0(x)=1$ is constant (for $x=0..n-1$), the eigenvalue $\lambda_0$ of $[\mathbf{A}]$ for the eigenvector $|\mathbf{U}_0\rangle$ is:

$$\lambda_0 = m \geq \lambda_k \tag{4.12}$$

From eq. (4.11) it also follows that $\lambda_0 \geq \lambda_k$ for $k=1..n-1$ since the sum in eq. (4.11) may contain one or more negative addends $\mathbf{U}_k(h_s)=-1$ for $k>0$, while for the k=0 case all addends are equal to +1.

The results above generalize the solution of the eigenproblem given in [23] for regular hypercube to the most general $\mathbf{Z}_2^d$ based Cayley graph.

## B. Computing Bisection

### 1) Cuts from adjacency matrix and partition vector

By definition [from section 3](), bisection **B** is computed by finding the minimum cut **C(X)** in the set **E**={**X**} of all possible equipartitions **X**=S$_1$+S$_2$ of the set of *n* vertices. An equipartition **X** can be represented by an *n*-dimensional vector |**X**⟩ ∈ $\mathbb{V}_n$ containing *n*/2 values +1 selecting nodes of group S$_1$, and *n*/2 values -1 selecting the nodes of group S$_2$. Since the cut value of a given equipartition **X** does not depend on particular +1/-1 labeling convention (e.g. changing sign of all elements $x_i$ defines the same graph partition), all vectors |**X**⟩ will have by convention the 1$^{st}$ component set to 1 and only the remaining *n*-1 components need to be varied (permuted) to obtain all possible distinct equipartitions from **E**. Hence, the equipartitions set **E** consists of all vectors **X** = ($x_0, x_1, \ldots x_{n-1}$), where $x_0$=1, $x_i \in \{+1,-1\}$ and $\sum_{i=0}^{n-1} x_i = 0$.

The cut value **C(X)** for a given partition **X** = ($x_0, x_1, \ldots x_{n-1}$) is obtained as the count of links which cross between nodes in S$_1$ and S$_2$. Such links can be easily identified via **E** and adjacency matrix [**A**], since [**A**]$_{i,j}$ is 1 iff nodes *i* and *j* are connected and 0 if they are not connected. The group membership of some node *i* is stored in the component $x_i$ of the partition **X**. Therefore, the links (*i, j*) that are counted have [**A**]$_{i,j}$=1, i.e. nodes *i* and *j* must be connected, and they must be in opposite partitions i.e. $x_i \neq x_j$. Recalling that $x_i$ and $x_j$ have values +1 or -1, the "$x_i \neq x_j$" is equivalent to "($x_i \cdot x_j$)=-1". To express that condition as a contribution +1 when $x_i \neq x_j$ and a contribution 0 when $x_i = x_j$, expression (1- $x_i \cdot x_j$)/2 is constructed which yields precisely the desired contributions +1 and 0 for any $x_i, x_j = \pm 1$. Hence, the values added to the link count can be written as **C**$_{i,j} \equiv$ (1- $x_i \cdot x_j$)·[**A**]$_{i,j}$/2 since **C**$_{i,j}$=1 iff nodes *i* and *j* are connected ([**A**]$_{i,j}$=1) *and* they are in different groups ($x_i \cdot x_j$=-1). Otherwise **C**$_{i,j}$ is 0, thus adding no contribution to the **C(X)**.

A counting detail that needs a bit of care arises when adding **C**$_{i,j}$ terms for *all i,j*=0..*n*-1. Namely, if the contribution of e.g. **C**$_{3,5}$ for nodes 3 and 5 is 1, because [**A**]$_{3,5}$=1 (3,5 linked), $x_3$=-1 and $x_5$=+1, then the contribution of the same link will contribute also via **C**$_{5,3}$ term since [**A**]$_{5,3}$=1, $x_5$=+1, $x_3$=-1. Hence the sum of **C**$_{i,j}$ for all *i,j*=0..*n*-1 counts the contribution for each link twice. Therefore, to compute the cut value **C(X)** for some partition **X**, the sum of **C**$_{i,j}$ terms must be divided by 2. Noting also that for any vector **X**∈**E** ⇒ ⟨**X**|**X**⟩ = $\sum_{i=0}^{n-1} x_i x_i = n$ and $\sum_{i,j=0}^{n-1} [A]_{i,j} = \sum_{j=0}^{n-1} m = n \cdot m$, yields for the cut **C(X)**:

$$\mathbf{C}(\mathbf{X}) = \frac{1}{2} \sum_{i,j=0}^{n-1} \frac{1}{2} (1 - x_i x_j) [A]_{i,j} = \frac{nm}{4} - \frac{1}{4} \sum_{i,j=0}^{n-1} x_i x_j [A]_{i,j} = \frac{n}{4}\left(m - \frac{\langle \mathbf{X}|\mathbf{A}|\mathbf{X}\rangle}{\langle \mathbf{X}|\mathbf{X}\rangle}\right) \quad (4.14)$$

To illustrate operation of the formula (4.14), the Fig. 4.5 shows adjacency matrix [**A**] for ***Cay***($\mathbf{Z}_2^4$,**S**$_5$), which reproduces $\mathbb{FQ}_4$ (folded 4-cube), with *d*=4, *n*=2$^d$=2$^4$=16 nodes, *m*=5 links per node, produced by the generator set **S**$_5$={1, 2, 4, 8, F}$_{hex}$={0001, 0010, 0100, 1000, 1111}$_{bin}$. The row and column headers show the sign pattern of the example partition **X**=(1,1,1,1, -1,-1,-1,-1, 1,1,1,1, -1,-1,-1,-1) and the shaded areas indicate the blocks of [**A**] in which eq. (4.14) counts ones – elements of [**A**] where row *r* and column *c* have opposite signs of the **X** components $x_r$ and $x_c$. The cut is computed as **C(X)**= ½ (sum of ones in shaded blocks) = 1/2*(4*8) = 16 which is the correct **B** for $\mathbb{FQ}_4$ (2*$n$/2=2*8=16). Note that the zeros (they don't contribute to **C(X)**) in the matrix [**A**] are shown as '-' symbol.

| X | + | + | + | + | - | - | - | - | + | + | + | + | - | - | - | - |
|---|---|---|---|---|---|---|---|---|---|---|---|---|---|---|---|---|
| + | - | 1 | 1 | - | 1 | - | - | - | 1 | - | - | - | - | - | - | 1 |
| + | 1 | - | - | 1 | - | 1 | - | - | - | 1 | - | - | - | - | 1 | - |
| + | 1 | - | - | 1 | - | - | 1 | - | - | - | 1 | - | - | 1 | - | - |
| + | - | 1 | 1 | - | - | - | - | 1 | - | - | - | 1 | 1 | - | - | - |
| - | 1 | - | - | - | - | 1 | 1 | - | - | - | - | 1 | 1 | - | - | - |
| - | - | 1 | - | - | 1 | - | - | 1 | - | - | 1 | - | - | 1 | - | - |
| - | - | - | 1 | - | 1 | - | - | 1 | - | 1 | - | - | - | - | 1 | - |
| - | - | - | - | 1 | - | 1 | 1 | - | 1 | - | - | - | - | - | - | 1 |
| + | 1 | - | - | - | - | - | - | 1 | - | 1 | 1 | - | 1 | - | - | - |
| + | - | 1 | - | - | - | - | 1 | - | 1 | - | - | 1 | - | 1 | - | - |
| + | - | - | 1 | - | - | 1 | - | - | 1 | - | - | 1 | - | - | 1 | - |
| + | - | - | - | 1 | 1 | - | - | - | - | 1 | 1 | - | - | - | - | 1 |
| - | - | - | - | 1 | 1 | - | - | - | 1 | - | - | - | - | 1 | 1 | - |
| - | - | - | 1 | - | - | 1 | - | - | - | 1 | - | - | 1 | - | - | 1 |
| - | - | 1 | - | - | - | - | 1 | - | - | - | 1 | - | 1 | - | - | 1 |
| - | 1 | - | - | - | - | - | - | 1 | - | - | - | 1 | - | 1 | 1 | - |

**Fig. 4.5**

## 2) Finding the minimum cut (bisection)

Bisection **B** is computed as the minimum cut **C(X)** for all **X**∈**E**, which via eq. (4.14) yields:

$$\mathbf{B} = \min_{\mathbf{X} \in \mathbf{E}} \left\{ \frac{n}{4} \left( m - \frac{\langle \mathbf{X}|\mathbf{A}|\mathbf{X} \rangle}{\langle \mathbf{X}|\mathbf{X} \rangle} \right) \right\} = \frac{nm}{4} - \frac{n}{4} \max_{\mathbf{X} \in \mathbf{E}} \left\{ \frac{\langle \mathbf{X}|\mathbf{A}|\mathbf{X} \rangle}{\langle \mathbf{X}|\mathbf{X} \rangle} \right\} \equiv \frac{nm}{4} - \frac{n}{4} \mathbf{M}_E \qquad (4.15)$$

$$where: \quad \mathbf{M}_E \equiv \max_{\mathbf{X} \in \mathbf{E}} \left\{ \frac{\langle \mathbf{X}|\mathbf{A}|\mathbf{X} \rangle}{\langle \mathbf{X}|\mathbf{X} \rangle} \right\} \qquad (4.16)$$

Despite the apparent similarity between the max{} term $\mathbf{M}_E$ in eq. (4.16) to the max{} term $\mathbf{M}_V$ in eq. (2.46), the Rayleigh-Ritz eqs. (2.45)-(2.46) do not directly apply to min{} and max{} expressions in eq. (4.15). Namely, the latter extrema are *constrained* to the set **E** of equipartitions, which is a *proper* subset of the full vector space $\mathbb{V}_n$ to which the Rayleigh-Ritz applies. The $\mathbf{M}_E \equiv$ max{} in eq. (4.16) can be smaller than the $\mathbf{M}_V \equiv$ max{} computed by eq. (2.46) since the result $\mathbf{M}_V$ can be a vector from $\mathbb{V}_n$ which doesn't belong to **E** (the set containing only the equipartition vectors **X**) i.e. if $\mathbf{M}_V$ is solved only by some vectors **Y** which do not consist of exactly $n$/2 elements +1 and $n$/2 elements -1.

As an illustration of the problem, $M_E$ is analogous to the "tallest programmer in the world" while $M_V$ is analogous to the "tallest person in the world." Since the set of "all persons in the world" (analogous to $\mathbb{V}_n$) includes as a *proper* subset the set of "all programmers in the world" (analogous to **E**) the tallest programmer may be shorter than the tallest person (e.g. the latter might be a non-programmer). Hence in general case the relation between the two extrema is $M_E \leq M_V$. The equality holds only if at least one solution from $M_V$ belongs also to **E**, or in the analogy, if at least one person among the "tallest person in the world" is also a programmer. Otherwise, strict inequality holds $M_E < M_V$.

In order to evaluate $M_E \equiv \max\{\}$ in eq. (4.16), the *n*-dimensional vector space $\mathbb{V}_n$ (the space to which vectors $|X\rangle$ belong) is decomposed into a [direct sum](#) of two mutually orthogonal subspaces:

$$\mathbb{V}_n = \mathbb{V}_0 \oplus \mathbb{V}_E \tag{4.17}$$

Subspace $\mathbb{V}_0$ is one dimensional space spanned by a single 'vector of all ones' $\langle 1|$ defined as:

$$\langle 1| \equiv (1,1,1,\ldots,1) \tag{4.18}$$

while $\mathbb{V}_E$ is the (*n*-1) dimensional orthogonal complement of $\mathbb{V}_0$ within $\mathbb{V}_n$, i.e. $\mathbb{V}_E$ is spanned by some basis of *n*-1 vectors which are orthogonal to $\langle 1|$. Using the eq. (2.6) for Walsh function $U_0(x)$, it follows:

$$\langle 1| \equiv (1,1,1,\ldots,1) = \langle U_0| \tag{4.19}$$

Hence, $\mathbb{V}_E$ is spanned by the remaining orthogonal set of *n*-1 Walsh functions $|U_k\rangle$, *k*=1..*n*-1. For convenience the latter subset of Walsh functions is labeled as set $\Phi$ below:

$$\Phi \equiv \{|U_k\rangle: \; k=1..n\text{-}1\} \tag{4.20}$$

Since all vectors $X \in E$ contain $n/2$ components equal +1 and $n/2$ components equal -1, then via (4.18):

$$\langle 1|X\rangle = \sum_{i=0}^{n-1} 1 \cdot x_i = 0, \;\; \forall X \in E \tag{4.21}$$

i.e. $\langle 1|$ is orthogonal to all equipartion vectors **X** from **E**, hence the entire set **E** is a proper subset of $\mathbb{V}_E$ (which is the set of all vectors $\in \mathbb{V}_n$ orthogonal to $\langle 1|$). Using $M_E$ in eq. (4.16) and eq. (2.46) results in:

$$M_E \equiv \max_{X \in E}\left\{\frac{\langle X|A|X\rangle}{\langle X|X\rangle}\right\} \leq M_V \equiv \max_{X \in \mathbb{V}_E}\left\{\frac{\langle X|A|X\rangle}{\langle X|X\rangle}\right\} = \lambda_{max} \tag{4.22}$$

The $M_V$ in eq. (4.22) is solved by an eigenvector $|Y\rangle$ of [A] for which $[A]|Y\rangle = \lambda_{max}|Y\rangle$ since:

$$\frac{\langle Y|A|Y\rangle}{\langle Y|Y\rangle} = \frac{\langle Y|\lambda_{max}|Y\rangle}{\langle Y|Y\rangle} = \frac{\lambda_{max}\langle Y|Y\rangle}{\langle Y|Y\rangle} = \lambda_{max} \tag{4.23}$$

Recalling, via eq. (4.10), that the eigenbasis of the adjacency matrix [A] in eq. (4.22) is the set of Walsh functions $|U_k\rangle$, and that $\mathbb{V}_E$ in which the $M_V=\max\{\}$ is searched for, is spanned by the *n*-1 Walsh functions $|U_k\rangle \in \Phi$, it follows that the eigenvector $|Y\rangle$ of [A] in eq. (4.23) can be selected to be one of these *n*-1 Walsh functions from $\Phi$ (since they form a complete eigenbasis of [A] in $\mathbb{V}_E$) i.e.:

$$|Y\rangle \in \Phi \equiv \{|U_k\rangle: \; k=1..n\text{-}1\} \tag{4.24}$$

The equality in (4.22) holds iff at least one solution $|Y\rangle \in \mathbb{V}_E$ is also a vector from the set **E**. In terms of the earlier analogy, this can be stated as: in the statement "the tallest student" ≤ "the tallest person", the equality holds iff at least one among the "tallest person" happens to be a "programmer."

Since $|Y\rangle$ is one of the Walsh functions from Φ and since all $|U_k\rangle \in \Phi$ have, via eqs. (2.5) and (2.7), exactly $n/2$ components equal +1 and $n/2$ components equal -1, $|Y\rangle$ belongs to the set **E**. Hence the exact solution for $M_E$ in eq. (4.22) is the Walsh functions $|U_k\rangle \in \Phi$ with the largest eigenvalue $\lambda_k$. Returning to the original bisection eq. (4.15), where $M_E$ is the second term, it follows that **B** is solved exactly by this same solution $|Y\rangle=|U_k\rangle \in \Phi$. Combining thus eq. (4.15) with equality case for $M_E$ in eq. (4.22) yields:

$$\mathbf{B} = \frac{nm}{4} - \frac{n}{4}\mathbf{M}_E = \frac{n}{4}(m - \lambda_{max}) = \frac{n}{4}\left(m - \max_{k\in[1,n)}\{\lambda_k\}\right) \tag{4.25}$$

Therefore, the computation of **B** is reduced to evaluating $n$-1 eigenvalues $\lambda_k$ of [A] for $k=1..n$-1 and finding a $t \equiv$ ($k$ with the largest $\lambda_k$) i.e. a $t$ such that $\lambda_t \geq \lambda_k$ for $k=1..n$-1. The corresponding Walsh function $U_t$ provides the equipartition which achieves this bisection **B** (the exact minimum cut). The evaluation of $\lambda_k$ in eq. (4.25) can be written in terms of the $m$ generators $h_s \in S_m$ via eq. (4.11) as:

$$\mathbf{B} = \frac{n}{4}\left(m - \max_{k\in[1,n)}\left\{\sum_{s=1}^{m}\mathbf{U}_k(h_s)\right\}\right) \tag{4.26}$$

Although the function values $U_k(x)$ above can be computed via eq. (2.5) as $\mathbf{U}_k(x) = (-1)^{\mathbb{P}(k\&x)}$, due to parallelism of binary operation on a regular CPU, it is computationally more efficient to use binary form of Walsh functions, $W_k(x)$. The binary ↔ algebraic translations in eqs. (2.8) can be rewritten in vector form for $U_k$ and $W_k$, with aid of definition of $|1\rangle$ from eq. (4.18), as:

$$|\mathbf{W}_k\rangle \equiv \tfrac{1}{2}(|1\rangle - |\mathbf{U}_k\rangle) \tag{4.27}$$

$$|\mathbf{U}_k\rangle = |1\rangle - 2\cdot|\mathbf{W}_k\rangle \tag{4.28}$$

Hence, the **B** formula (4.26) can be written in terms of $W_k$ via eq. (4.28) and $W_k$ formula eq. (2.10) as:

$$\mathbf{B} = \frac{n}{4}\left(m - \max_{k\in[1,n)}\left\{\sum_{s=1}^{m}(1 - 2\cdot\mathbf{W}_k(h_s))\right\}\right) = \frac{n}{4}\left(m - \max_{k\in[1,n)}\left\{m - 2\sum_{s=1}^{m}\mathbf{W}_k(h_s)\right\}\right) \Rightarrow$$

$$\Rightarrow \mathbf{B} = \frac{n}{2}\min_{k\in[1,n)}\left\{\sum_{s=1}^{m}\mathbf{W}_k(h_s)\right\} = \frac{n}{2}\min_{k\in[1,n)}\left\{\sum_{s=1}^{m}\mathbb{P}(k\&h_s)\right\} \tag{4.29}$$

The final expression in (4.29) is particularly convenient since for each $k=1..n$-1 it merely adds parities of the bitwise AND terms: ($k\&h_s$) for all $m$ Cayley graph generators $h_s \in S_m$. The parity function $\mathbb{P}(x)$ in eq. (4.29) can be computed efficiently via a short C function ([14] p. 42) as follows:

```
inline int Parity(unsigned int x)                                                        (4.30)
{
   x^=x>>16, x^=x>>8, x^=x>>4, x^=x>>2;
   return (x^(x>>1))&1;
}
```

Using a $\mathbb{P}(x)$ implementation **Parity(x)**, the entire computation of **B** via eq. (4.29) can be done by a small C function **Bisection(n,hops[],m)** as shown in code (4.31).

```
int Bisection(int n,int *ha,int m)                                                       (4.31)
{
  int cut,b,i,k;                  // n=2^d is # of nodes, m=# of hops

     for(b=n,k=1; k<n; ++k)       // Loop through all n-1 Wk() functions
       {                          // Set initial min cut b=n (out of range since m<n)
        for(cut=i=0; i<m; ++i)    // Loop through all m hops ha[i], add
          cut+=Parity(ha[i]&k);   // +1 if hop[i] coincides with Wk(hop[i])
        if (cut<b) b=cut;         // Update min cut if count cut<old_min_cut
       }
     return b;                    // Return bisection (min cut) in units n/2
}
```

The inner loop in (4.31) executes $m$ times and the outer loop ($n$-1) times, yielding total of ~ $m \cdot n$ steps of $\log(n)$ complexity each (for parity). Hence, the complexity of **B** is $O(m \cdot n \cdot \log(n))$. For large $m$ this may be further optimized using Walsh transform. First, we define function $f(x)$ for $x=0,1,\ldots n$-1 as:

$$f(x) \equiv [x \in \mathbf{S}_m] = \begin{cases} 1 & if \ x \in \mathbf{S}_m \\ 0 & if \ x \notin \mathbf{S}_m \end{cases} \tag{4.32}$$

where and $0 \leq x < n$ and $\mathbf{S}_m = \{h_1, h_2, \ldots h_m\}$ is the set of $m$ graph generators. Hence, $f(x)$ is 1 when $x$ is equal to one of the generators $h_s \in \mathbf{S}_m$ and 0 elsewhere. This function can be viewed as a vector $|f\rangle$, with components $f_i = f(i)$. Recalling the computation of adjacency matrix [**A**] via eq. (4.2), vector $|f\rangle$ can also be recognized as the 0-th column of [**A**] i.e. $f_i = [\mathbf{A}]_{0,i}$. With this notation, the eq. (4.26) for **B** becomes:

$$\mathbf{B} = \frac{n}{4}\left(m - \max_{k \in [1,n)}\left\{\sum_{s=1}^{m} \mathbf{U}_k(h_s)\right\}\right) = \frac{n}{4}\left(m - \max_{k \in [1,n)}\{\langle \mathbf{U}_k | f \rangle\}\right) \equiv \frac{n}{4}\left(m - \max_{k \in [1,n)}\{F_k\}\right) \tag{4.33}$$

$$where: \ F_k \equiv \langle \mathbf{U}_k | f \rangle \tag{4.34}$$

Therefore, the **B** computation consists of finding the largest element in the set $\{F_k\}$ of $n$-1 elements. Using the orthogonality and completeness of the $n$ vectors $|\mathbf{U}_k\rangle$, $\langle \mathbf{U}_j | \mathbf{U}_k \rangle = n \cdot \delta_{j,k}$ from eq. (2.3), important property of the set $\{F_k\}$ follows:

$$\sum_{k=0}^{n-1} \frac{1}{n} F_k |\mathbf{U}_k\rangle = \sum_{k=0}^{n-1} \frac{1}{n} |\mathbf{U}_k\rangle\langle \mathbf{U}_k | f \rangle = \left(\frac{1}{n}\sum_{k=0}^{n-1} |\mathbf{U}_k\rangle\langle \mathbf{U}_k|\right) |f\rangle = \mathbf{I}_n |f\rangle = |f\rangle \tag{4.35}$$

The eqs. (4.34),(4.35) can be recognized as the Walsh transform ([14] chap. 23) of function $f(x)$, with $n$ coefficients $F_k/n$ as the transform coefficients. Hence, computational complexity for $n$ coefficients $F_k$ in (4.34) is via Fast Walsh Transform $O(n \cdot \log(n))$, hence complexity for **B** in (4.33) is $O(n \cdot \log(n))$.

## C. Optimizing Bisection

### 1) Direct Optimization

With the obtained $O(n \cdot \log(n))$ complexity method for exact computation of bisection **B** for a given set of generators $\mathbf{S}_m$, the next task identified is the optimization of the generator set $\mathbf{S}_m = \{\boldsymbol{h}_1, \boldsymbol{h}_2, \ldots \boldsymbol{h}_m\}$ i.e. the finding of the $\mathbf{S}_m$ with the largest **B**. The individual hops $\boldsymbol{h}_s$ are are distinct and constrained to $n$-1 values: 1,2,… $n$-1 (0 is excluded since no node is connected to itself), i.e. $\mathbf{S}_m$ is an **m** element subset of the integer sequence 1..$n$-1. For convenience, this set of all **m**-subsets of integer sequence 1..$n$-1 is labeled as:

$$\Omega(n,m) \equiv \Omega_n \equiv \{\mathbf{S}_m : (\mathbf{S}_m = \{h_1, h_2, \ldots, h_m\}) \text{ and } (0 < h_s < n) \text{ and } (h_s \neq h_r)\} \quad (4.40)$$

$$|\Omega| \equiv |\Omega(n,m)| = \binom{n-1}{m} = O(n^m) \quad (4.41)$$

With this notation and using the binary formula for **B**, eq. (4.29), the **B** optimization task is:

$$\mathbf{b} \equiv \frac{\mathbf{B}}{n/2} = \max_{\mathbf{S}_m \in \Omega_n} \left\{ \min_{k \in [1,n)} \left\{ \sum_{s=1}^{m} \mathbf{W}_k(h_s) \right\} \right\} \quad (4.42)$$

For convenience, eq. (4.42) also defines a quantity **b** which is the **bisection in units $n$/2**. The worst case computational complexity the **B** optimization is thus $O((m \cdot n \cdot \log(n))^m)^*$, which is polynomial in $n$, hence, at least in principle, it is a computationally tractable problem as $n$ increases. Note that **m** is typically a hardware characteristics of the network components, such as switches, which usually don't get replaced often as network size $n$ increases.

Since for large enough $n$, even a low power polynomial can render 'an in principle tractable' problem practically intractable, approximate methods for the max{} part of the computation (4.42) would be used in practice. Particularly attractive for this purpose would be genetic algorithms and simulated annealing techniques used in [12] (albeit for the task of computing **B**, which the methods of (2.B) solve efficiently and exactly). Some of the earlier implementations of LH construction have used fast greedy algorithms, which work fairly well. The optimization technique described next does not perform any such direct optimization of eq. (4.42), but uses a far more effective approach instead.

---

[*] The actual exponent here would be ($m$ - log($n$) – 1), not **m**, since the Cayley graphs are highly symmetrical and one would not have to search over the symmetrically equivalent subsets $\mathbf{S}_m$.

## 2) Bisection optimization via EC Codes

In order to describe this method, the inner-most term within the nested **max{min{}}** expression in the eq. (4.42) is identified and examined in more detail. For convenience, this term, which has a meaning of a cut for a partition defined via the pattern of ones in the Walsh function $W_k(x)$, is labeled as:

$$C_k \equiv \sum_{s=1}^{m} W_k(h_s) = \sum_{s=1}^{m} \mathbb{P}(k \& h_s) \quad (4.43)$$

Eq. (4.43) also expresses $W_k(x)$ in terms of parity function $\mathbb{P}(x)$ via eq. (2.10). The function $\mathbb{P}(x)$ for some $d$-bit integer $x = (x_{d-1} \ldots x_1 x_0)_{binary}$ is defined as:

$$\mathbb{P}(x) \equiv \left( \sum_{\mu=0}^{d-1} x_\mu \right) \bmod 2 = (x_0 \wedge x_1 \wedge \cdots \wedge x_{d-1}) \quad (4.44)$$

The last expression in eq. (4.44) shows that $\mathbb{P}(x) \equiv \mathbb{P}(x_{d-1} \ldots x_1 x_0)$ is a "linear combination" in terms of the selected field $\mathbf{GF}(2)^d$, of the field elements provided in the argument. The eq. (4.43) contains a modified argument of type $\mathbb{P}(k \& h)$, for $h \in S_m$, which can be reinterpreted as: the 'ones' from the integer $k$ are selecting a subset of bits from the $d$-bit integer $h$, then $\mathbb{P}(x)$ performs the linear combination of the selected subset of bits of $h$. For example, if $k = 11_{dec} = 1011_{bin}$ than the *action of* $W_{1011}(h) \equiv \mathbb{P}(1011 \& h)$ is to compute linear combination of the bits bit-0,1 and 3 of $h$ (bit numbering is zero based, from low/right to high/left significance). Since eq. (4.43) performs the above "linear combination via ones in $k$" *action* of $W_k$ on a series of $d$-bit integers $h_s$, $s=1..m$, the $W_k$ "action" on such series of integers is interpreted as the parallel linear combination on the bit-columns of the list of $h_s$ as shown in the Fig 4.6, for $k=1011$ and $W_{1011}$ acting on a set of generators $S_5 = \{ 0001, 0010, 0100, 1101 \}$. The 3 bit-columns $V_3$, $V_1$ and $V_0$ selected by ones in $k$ are combined via XOR into the resulting bit-column $V$: $|V_3\rangle \oplus |V_1\rangle \oplus |V_0\rangle = |V\rangle$.

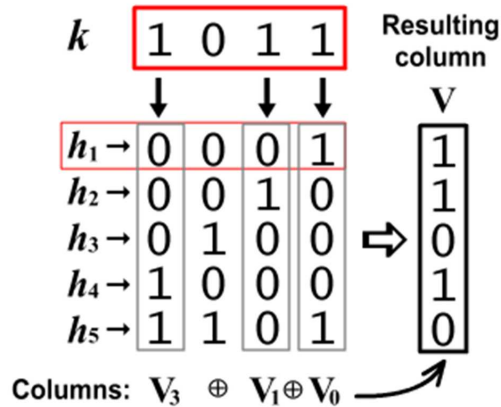

**Fig. 4.6**

Therefore, the action of a $\mathbf{W}_k$ on the generator set $\mathbf{S}_m = \{h_1, h_2, \ldots h_m\}$ can be seen as a "linear combination" of the length-$m$ columns of digits (columns selected by ones in $k$ from $\mathbf{W}_k$) formed by the $m$ generators $h_s$. If instead of the $\mathbf{Z}_2^d$ used in the example of Fig. 4.6, there was a more general Cayley graph group, such as $\mathbf{Z}_q^d$, instead of the bit-columns there would have been length-$m$ columns made of digits in alphabet of size $q$ (i.e. integers $0..q$-1) and the XOR would have been replaced with the appropriate $\mathbf{GF}(q)$ field arithmetic e.g. addition modulo $q$ on $m$-tuples for $\mathbf{Z}_q^d$ as illustrated in an earlier example in Fig. 4.1. The construction of column vectors $|\mathbf{V}_\mu\rangle$ of Fig. 4.6 can be expressed more precisely via an $m \times d$ matrix $[\mathbf{R}_{m,d}]$ defined as:

$$[\mathbf{R}_{m,d}] \equiv \sum_{s=1}^{m} |e_s\rangle\langle h_s| = \begin{pmatrix} \langle h_1| \\ \ldots \\ \langle h_m| \end{pmatrix} = \begin{pmatrix} h_{1,d-1} & h_{1,d-2} & \ldots & h_{1,0} \\ \ldots & \ldots & \ldots & \ldots \\ h_{m,d-1} & h_{m,d-2} & \ldots & h_{m,0} \end{pmatrix} \equiv (|\mathbf{V}_{d-1}\rangle, |\mathbf{V}_{d-2}\rangle, \ldots |\mathbf{V}_0\rangle) \quad (4.45)$$

$$\text{where: } \left(|\mathbf{V}_\mu\rangle\right)_s \equiv h_{s,\mu} = (\langle h_s|)_\mu \quad \text{for } \mu = 0..d-1, s = 1..m \quad (4.46)$$

Hence the $m$ rows of matrix $[\mathbf{R}_{m,d}]$ are $m$ generators $\langle h_s| \in \mathbf{S}_m$ and its $d$ columns are $d$ column vectors $|\mathbf{V}_\mu\rangle$. The above 'linear combination of columns via ones in $k$' becomes in this notation:

$$|\mathbf{V}(k)\rangle \equiv \sum_{\mu=0}^{d-1} k_\mu |\mathbf{V}_\mu\rangle \quad \text{where } k \equiv \sum_{\mu=0}^{d-1} k_\mu 2^\mu \quad (4.47)$$

where the linear combination of $k_\mu|\mathbf{V}_\mu\rangle$ is performed in $\mathbf{GF}(q)$ i.e. mod $q$ on each component of $m$-tuples $k_\mu|\mathbf{V}_\mu\rangle$. The sum computing the cut $\mathbf{C}_k$ in eq. (4.43) is then simply adding (*without* mod $q$) all components of the vector $|\mathbf{V}(k)\rangle$ from eq. (4.47). Recalling the definition of Hamming weight as the number of non-zero digits, this cut $\mathbf{C}_k$ is recognizable as the Hamming weight of the vector $|\mathbf{V}(k)\rangle$:

$$\mathbf{C}_k = \langle \mathbf{V}(k) \rangle \quad (4.48)$$

The next step is to propagate the new "linear combination" interpretation of $\mathbf{W}_k$ action back one more level, to the original optimization problem in eq. (4.42), in which the cut $\mathbf{C}_k$ was only the innermost term. The min{} block of eq. (4.42), seeks a minimum value of $\mathbf{C}_k$ for all $k=1..n$-1. The set of $n$ vectors $|\mathbf{V}(k)\rangle$ obtained via eq. (4.47) when $k$ runs through all possible integers $0..n$-1 is a $d$-dimensional vector space, a linear span (subspace of $m$-tuples vector space $\mathbb{V}_m$), which is denoted as $\mathbb{S}(d,m,q)$:

$$\mathbb{S}(d,m,q) \equiv \{|\mathbf{V}(k)\rangle : k = 0..n-1\} \quad (4.49)$$

Therefore, the min{} level optimization in eq. (4.42) computing bisection $\mathbf{b}$, seeks a non-zero vector $|\mathbf{V}(k)\rangle$ from the linear span $\mathbb{S}(d,m,q)$ with the smallest Hamming[*] weight $\langle \mathbf{V}(k) \rangle$:

$$\mathbf{b} = \min\{\langle \mathbf{V}(k) \rangle : (\mathbf{V}(k) \in \mathbb{S}(d,m,q)) \text{ and } (\mathbf{V}(k) \neq 0)\} \quad (4.50)$$

But $\mathbf{b}$ in eq. (4.50) is precisely the definition eq. (2.25) of the minimum weight $w_{min}$ in the codeword space (linear span) $\mathbb{S}(\_k,\_n,q)$ of non-zero codewords $\mathbf{Y}$. Note: In order to avoid the mix up in the notation between the two fields, the overlapping symbols [*n*, *k*] which have a different meaning in ECC, will in this section have an underscore prefix, i.e. the linear code [*n*, *k*] is relabeled as [_*n*, _*k*].

---

[*] Hamming weight was used in the current LH implementation, but any other weight, such as Lee weight, could have been used, which would correspond to other Cayley graph groups $\mathbf{G}_n$ and generator sets $\mathbf{S}_m$.

The mapping between the ECC quantities and LH quantities is then: $w_{min} \Leftrightarrow b$, _*k* $\Leftrightarrow d$, _*n* $\Leftrightarrow m$, _*k* vectors $\langle g_i|$ spanning linear space $\mathbb{S}(\_k,\_n,q)$ of _*n*-tuples and constructing code generator matrix [G] (eq. (2.20)) $\Leftrightarrow d$ columns $|V_\mu\rangle$ for µ=0..*d*-1 spanning linear space $\mathbb{S}(d,m,q)$ of *m*-tuples (digit-columns in the generator list). Since, via eq. (2.26) the minimum weight of the code $w_{min}$ is same as the minimum distance Δ between the codewords **Y**, it follows that the bisection **b** is also the same quantity as the ECC Δ (even numerically). The table in Fig. 4.7 lists the most important elements of this mapping.

| Linear EC codes | _*k* | _*n* | $\mathbb{S}(\_k,\_n,q)$ | Δ | _*k* rows of _*n*-tuples $\langle g_i|$ | *q* for **GF**(*q*) |
|---|---|---|---|---|---|---|
| LH Networks | *d* | *m* | $\mathbb{S}(d,m,q)$ | **b** | *d* columns of *m*-tuples $|V_\mu\rangle$ | *q* for $\mathbf{Z}_q^d$ |

**Fig. 4.7**

The optimization of linear code [_*n*, _*k*, Δ] that maximizes Δ is thus the same optimization as the outermost level of the LH optimization, max{} level in eq. (4.42) that seeks the Cayley graph generator set **S**$_m$ with the largest bisection **b** – other than difference in labeling conventions, both optimizations seek the *d*-dimensional subspace $\mathbb{S}(d,m,q)$ of some vectors space $\mathbb{V}_m$ which maximizes the minimum non-zero weight $w_{min} \Leftrightarrow b$ of the subspace $\mathbb{S}$. The two problems are mathematically one and the same.

Therefore, the vast numbers of good/optimal linear ECC codes computed over the last six decades (such as EC code tables [17] and [21]) are immediately available as good/optimal solutions for the **b** optimization problem of the LH networks, such as eq. (4.42) for Cayley graph group $\mathbf{G}_n = \mathbf{Z}_q^d$. Similarly any techniques, algorithms and computer programs (e.g. MAGMA ECC module) used for constructing and combining of good/optimum linear EC codes, such as quadratic residue codes, Goppa, Justesen, BCH, cyclic codes, Reed-Muller codes,… [15],[16], via translation table in Fig. 4.7, automatically become techniques and algorithms for constructing good/optimum LH networks.

As an illustration of the above translation procedure, a simple parity check EC code [4,3,1]$_2$ with generator matrix [**G**$_{3,4}$] is shown in Fig. 4.8. The codeword has 1 parity bit followed by 3 message bits and is capable of detecting all single bit errors. The translation to the optimum network shown on the right, is obtained by rotating 90° counter-clockwise ↺ the 3×4 generator matrix [**G**$_{3,4}$]. The obtained block of 4 rows with 3 bits per row is interpreted as 4 generators $h_s$, each 3 bits wide, for the $Cay(\mathbf{Z}_2^3, \mathbf{C}_4)$ graph. The resulting network thus has *d*=3, *n*=2$^3$=8 nodes and *m*=4 links/node. The actual network is a folded 3-cube shown within an earlier example in Fig. 4.4. Its bisection is: **b**=2 and **B**=**b**·*n*/2=8 links.

$$[\mathbf{G}_{3,4}] = \begin{pmatrix} 1 & \mathbf{1} & 0 & 0 \\ 1 & 0 & \mathbf{1} & 0 \\ 1 & 0 & 0 & \mathbf{1} \end{pmatrix} \implies \mathbf{C}_4 = \begin{matrix} h_1 = 001 = 1 \\ h_2 = 010 = 2 \\ h_3 = 100 = 4 \\ h_4 = 111 = 7 \end{matrix}$$

**Fig. 4.8**

A slightly larger and denser network using EC code [7,4,3]$_2$ from Fig. 2.4, is converted into an optimum solution, a graph $Cay(\mathbf{Z}_2^4, \mathbf{C}_7)$, with $d$=4, $n$=16 nodes and $m$=7 link/node as shown in Fig. 4.9.

$$[\mathbf{G}_{4,7}] = \begin{pmatrix} 1 & 1 & 0 & \mathbf{1} & 0 & 0 & 0 \\ 0 & 1 & 1 & 0 & \mathbf{1} & 0 & 0 \\ 1 & 1 & 1 & 0 & 0 & \mathbf{1} & 0 \\ 1 & 0 & 1 & 0 & 0 & 0 & \mathbf{1} \end{pmatrix} \Rightarrow \mathbf{C}_7 = \begin{matrix} h_1 = & 0001 = 1 \\ h_2 = & 0010 = 2 \\ h_3 = & 0100 = 4 \\ h_4 = & 1000 = 8 \\ h_5 = & 0111 = 7 \\ h_6 = & 1110 = E \\ h_7 = & 1011 = B \end{matrix}$$

**Fig. 4.9**

The 4 row, 7 column generator matrix [$\mathbf{G}_{4,7}$] of the linear EC code [7,4,3]$_2$ on the left side was rotated 90° counter-clockwise and the resulting 7 rows of 4 digits are binary values for the 7 generators $h_s$ (also shown in hex) of the 16 node Cayley graph. The resulting $n$=16 node network has relative bisection (in $n$/2 units) $b$=Δ=3 and absolute bisection (in # of links) of: $\mathbf{B} = b \cdot n/2 = 3 \cdot 16/2 = 24$ links. Since the network is a non-planar 4-dimensional cube with total $n \cdot m/2 = 16 \cdot 7/2 = 56$ links it is not drawn.

The above examples are captured by the following simple, direct translation recipe:

$$\text{EC code } [\_n, \_k, \Delta]_q \quad \rightarrow \quad \text{LH } Cay(\mathbf{Z}_q^d, \mathbf{S}_m) \tag{4.45}$$

(i) Take EC code generator matrix [$\mathbf{G}_{\_k,\_n}$] and rotate it 90° (in either direction[*])
(ii) The result is $m = \_n$ row by $d = \_k$ column matrix [$\mathbf{R}_{m,d}$] of $\mathbf{GF}(q)$-digits 0..$q$-1
(iii) Read $m$ rows of $d$-tuples in base $q$ from [$\mathbf{R}_{m,d}$] as $m$ generators $h_s \in \mathbf{S}_m \subset \mathbf{Z}_q^d$
(iv) <u>Compute Cayley graph</u> LH=$Cay(\mathbf{Z}_q^d, \mathbf{S}_m)$ from the obtained generators $\mathbf{S}_m = \{h_1, h_2,.. h_m\}$
(v) LH: $n$=$q^d$ nodes, $m$ links/node, bisection: relative $b$=Δ, absolute $\mathbf{B}$=Δ·$n$/2 links

---

[*] Direction of rotation merely selects order of generators in the list, which is an arbitrary convention.

## D) Implementation Notes

### N-1. Equivalent LH networks

Order of elements in a generator set $S_m = \{ h_1, h_2, \ldots h_m \}$ is clearly a matter of convention and network performance characteristics don't depend on a particular ordering. Similarly, the subspace $\mathbb{S}(d,m,q)$ of the column vectors can be generated using any linearly independent set of $d$ vectors from $\mathbb{S}(d,m,q)$ instead of the original subset $\{V_\mu\}$. All these transformation of a given network yield equivalent networks, differing only in labeling convention but all with the same distribution of cuts (including min-cut and max-cut) and the same network paths distribution (e.g. same average and max paths). This equivalence is used to compute specific generators optimized for some other objective, beyond the cuts and paths. Some of these other objectives are listed in the notes below.

### N-2. Minimum change network expansion

During expansion of the network, it is useful that the next larger network is produced with the **minimum change** from the previous configuration e.g. requiring the fewest cables to be reconnected to other switches or ports. The equivalence transforms of N-1 are used to "morph" the two configuration, initial and final toward each other, using the number of different links in $S_m$ as the cost function being minimized.

### N-3. Diagonalization

It is often useful, especially in physical wiring, discovery and routing, to have a $Z_q^d$ based network in which (usually first) $d$ hops from $S_m$ are powers of $q$. This property of generator set $S_m$ corresponds to [systematic generator](#) matrix $[G_{\_k,\_n}]$ for linear codes and can be recognized by the presence of identity matrix $I_d$ within $[G_{\_k,\_n}]$ (possibly with permuted columns). The two previous examples, Figures [4.8](#) and [4.9](#) were of this type (the digits of $I_d$ sub-matrix were in bold).

A simple, efficient method for computing a "systematic generator" from non-systematic one is to select for each column $c = 0..d-1$ a row $r(c)=1..m$ that contains a digit 1 in column $c$. If row $r(c)$ doesn't contain any other ones, then we have one column with desired property (the $h_{r(c)}$ is a power of 2). If there are any other columns, such as $c'$ which contain ones in row $r(c)$, the column $V_c$ is XOR-ed into these columns $V_{c'}$, clearing the excessive ones in $r(c)$. Finally, when there is a single 1 in row $r(c)$ and column $c$, the hop $h_{r(c)}$ is swapped with hop $h_{c+1}$ so that the resulting matrix contains generator $h_{c+1}=2^c$. The process is repeated for the remaining columns $c < d$.

The number of XOR operations between columns needed to reduce some row $r(c)$ to a single 1 in column $c$, is $\langle h_{r(c)} \rangle - 1$. Therefore, to reduce number of required XOR-s (columns are $m$ bits long which can be much larger than the machine word), for each new $c$ to diagonalize, algorithm picks the row which has the smallest weight, $\min\{\langle h_{r(c)} \rangle\}$.

## N-4. Digital or (t,m,s) nets (or designs, orthogonal arrays)

This research field is closely related to design of optimal linear codes [_*n*,_*k*,Δ]_q (cf. [20],[21]). The basic problem in the field of 'digital nets' is to find distribution of points on **s**-dimensional hypercubic (fish-) net with "binary intervals" layout of 'net eyes' (or generally analogous *b*-ary intervals via powers of any base *b*, not only for *b*=2) which places the same number of points into each net eye. There is a mapping between (*t*,_*m*,*s*)_b digital nets and [_*n*,_*k*]_q codes via identities: _*n*=*s*, _*k*=*s*-_*m*, *q*=*b*. A large database of optimal (*t*,_*m*,*s*) nets, which includes linear code translations is available via a web site [21]. Therefore, the solutions, algorithms and computer programs for constructing good/optimal (*t*,_*m*,*s*) nets are immediately portable to construction of good/optimal LH networks via this mapping followed by the [_*n*,_*k*]_q → LH mapping in Fig. 4.7.

## N-5. Non-binary codes

The linear codes with *q*>2 generate hyper-torus/-mesh type of networks of extent *q* when the Δ metrics of the code is Lee distance. When Hamming distance is used for *q*>2 codes, the networks are of generalized hypercube/flattened butterfly type [3]. For *q*=2, which is the binary code, the two types of distance metrics are one and the same.

## N-6. Non-binary Walsh functions

Walsh functions readily generalize to other groups, besides cyclic group $\mathbf{Z}_2^d$ used here (cf. [22]). A simple generalization to base *q*>2 for groups $\mathbf{Z}_q^d$, for any integer *q* is based on defining function values via *q*-th primitive root of unity ω:

$$\mathbf{U}_{q,k}(x) = \omega^{\sum_{\mu=0}^{d-1} k_\mu x_\mu} \; for \; x, k < n \equiv q^d \tag{4.50}$$

$$where: \quad \omega \equiv e^{2\pi i/q} \tag{4.51}$$

For *q*=2, eq. (4.51) yields ω=(-1), which reduces $\mathbf{U}_{q,k}(x)$ from eq. (4.50) to the regular Walsh functions $\mathbf{U}_k(x)$, eq. (2.5). The *q* discrete values of $\mathbf{U}_{q,k}(x)$ can also mapped into integers in [0,*q*) interval to obtain integer-valued Walsh functions $\mathbf{W}_{q,k}(x)$ (analogue of binary form $\mathbf{W}_k(x)$), useful for efficient computer implementation, via analogous mapping to the binary case e.g. via mapping $a = \omega^b$ for integer *b*=0..*n*-1, where *b*:integer, *a*:algebraic value, as in eq. (2.8) where this same mapping (expressed differently) was used for *q*=2.

The non-binary Walsh functions $\mathbf{U}_{q,k}$ can be used to define graph partition into *f* parts where *f* is any divisor of *q* (including *q*). For even *q*, this allows for efficient computation of bisection. The method is a direct generalization of the binary case: the *q* distinct function values of $\mathbf{U}_{q,k}(x)$ define partitions arrays $\mathbf{X}_k[x] \equiv \mathbf{U}_{q,k}(x)$ containing *n*=*q*^d elements indexed by *x*=0..*n*-1. Each of *q* values of $\mathbf{U}_{q,k}(x)$ indicates a node *x* belongs to one of the *q* parts. The partitions $\mathbf{X}_k$ for *k*=1..*n*-1 are examined and cuts computed using the adjacency matrix [**A**] for $Cay(\mathbf{Z}_q^d, \mathbf{S}_m)$ graph, as in eq. (4.14) for *q*=2. The generators **T**(*a*) and adjacency matrix [**A**] are computed via general eqs. (4.1),(4.2), where ⊕ operator is **GF**(*q*) addition (mod *q*).

## N-7. Secondary Optimizations

Once the optimum solution for (4.42) is obtained (via ECC, Digital nets, or via direct optimization), secondary optimizations, such as seeking the minimum diameter (max distance) or minimum average distance or largest max-cut, can be performed on the solution via local, greedy algorithms. Such algorithms were used in construction of our data solutions data base, where each set of parameters (*d*,*m*,*q*) has alternate solutions optimized for some other criteria (usually diameter, then average distance).

The basic algorithm attempts replacement of typically 1 or 2 generators[*] $h_s \in S_m$, and for each new configuration it evaluates (incrementally) the target utility function, such as diameter, average distance or max-cut (or some hierarchy of these, used for tie-breaking rules). The utility function also uses indirect measures (analogous to sub-goals) as a tie-breaking selection criterium e.g. when minimizing diameter, it was found that an effective indirect measure is the number of nodes #F in the farthest (from node 0) group of nodes. The indirect objective in this case would be to minimize the #F of such nodes, whenever the examined change (swap of 1 or two generators) leaves the diameter unchanged.

In addition to incremental updates to the networks after each evaluated generators replacement, these algorithms rely on vertex symmetry of Cayley graphs to further reduce computations. E.g. all distance tables are only maintained and updated for *n*-1 distances from node 0 ("root"), since the table is the same for all nodes (with mere permutation of indices, obtainable via **T**(*a*) representation of $G_n$ if needed).

Depending on network application, the bisection **b** can be maintained fixed for all replacements (e.g. if bisection is the highest valued objective), or one can allow **b** to drop by some value, if the secondary gains are sufficiently valuable.

After generating and evaluating all replacements to a given depth (e.g. replacement of 1 or 2 generators), the "best" one is picked (according to the utility/cost function) and replacement is performed. Then the outer iteration loop would continue, examining another set of replacements seeking the best one, etc. until no more improvements to the utility/cost function can be obtained in the last iteration.

## N-8. Asymmetrical Long Hop Networks (LH/A)

In practice one often needs a network which can expand in finer steps than in powers of 2 (or of prime *q*) available via the presented LH construction. While some additional size flexibility can be achieved by changing the Cayley graph group, e.g. to symmetric group $S_n$ yields network sizes *n*=*d*! for (d=2,3,..), known as Star graph [11] (corresponding to non-linear EC codes in our mapping), the available sizes are still much too sparse from the practical perspective. During exploration of various groups and truncation methods, a remarkable Cayley group was found[†] with a natural truncation to any size *n*, such that the necessary loss of CG symmetry is compensated by additional gains in network performance (throughput, max and average hops). These performance gains (10-40%) are the result of the vastly expanded solution space which became available due to relaxation of the CG symmetry contraints. The price paid for these gains are more complex forwarding, routing and wiring compared to the symmetrical LH networks.

---

[*] The number of simultaneous replacements depends on *n*, *m* and available computing resources. Namely, there are ~ $n^r$ possible simultaneous deletions and insertions (assuming the "best deletion" is followed by "best" insertion).
[†] These results will be presented in a separate, later paper.

# E) Specialized solution

This section describes several optimum LH solutions with particularly useful parameters or simple construction patterns.

**S-1. High Density LH Networks for modular switches (LH-HD)**

This is a special case of LH networks with high topological link density, suitable for combining smaller number of smaller radix switches into a single larger radix modular switch. This is a specialized domain of network parameters where the 2-layer Fat Tree (FT-2) networks are currently used since they achieve the yield of **E**=**R**/3 external ports/switch, which is the maximum mathematically possible for the worst case traffic patterns. The 'high density' LH networks (**LH-HD**) match the FT-2 in this optimum **E**=**R**/3 external ports/switch yield for the worst case traffic patterns, while achieving substantially lower average latency and the cost in Gb/s of throughput on random or 'benign' (non-worst case) traffic.

In our implementation using $Cay(\mathbf{Z}_2^d, \mathbf{S}_m)$ graph, the network size is $n=2^d$ switches *and* the number of links per node $m$ is one of the numbers: $n/2$, $n/2+n/4$, $n/2+n/4+n/8$,… , $n/2+n/4+n/8+…+1$, then the optimum $m$ generators for LH-HD are constructed as follows:

(i) $h_1=n-1, h_2=n-2, h_3=n-3,… h_m=n-m$
(ii) Optionally diagonalize and sort $\mathbf{S}_m$ via procedure (N-3)[*]

The resulting bisection is: $\mathbf{b}=\lfloor(m+1)/2\rfloor$ or $\mathbf{B}=\mathbf{b}\cdot n/2$, diameter is 2 and average hops is 2-$m/n$. The largest LH-HD $m = n/2+n/4+n/8+…+1 = n-1$ has $\mathbf{b}=n/2$ and corresponds to a fully meshed network.

Figure (4.10) shows an example of LH-HD generators for $n=2^6=64$ nodes and $m=n/2=32$ hops/node, with the hops shown in hex and binary (binary 0s are shown as '-' character). Fig. (4.10-a) shows the non-diagonalized hops after the step (i), and Fig. (4.10-b) shows the equivalent network with $m=32$ hops after diagonalization in step (ii) and sorting. Other possible LH-HD $m$ values for the same $n=64$ node network are $m=32+16=\mathbf{48}$, $m=48+8=\mathbf{56}$, $m=56+4=\mathbf{60}$, $m=60+2=\mathbf{62}$ and $m=61+1=63$ hops.

Additional **modified LH-HD networks** are obtained from any of the above LH-HD networks via removal of any one or two generators, which yields networks LH-HD1 with $m_1 = m$-1 and LH-HD2 with $m_2=m$-2 generators. Their respective bisections are $\mathbf{b}_1=\mathbf{b}$-1 and $\mathbf{b}_2=\mathbf{b}$-2. These two modified networks may be useful when an additional one or two server ports are needed on each switch compared to the unmodified LH-HD network.

These three types of high density LH networks are useful for building modular switches, networks on a chip in multi-core or multi-processor systems, flash memory/storage network designs, or generally any of the applications requiring very high bisection from a small number of high radix components and where FT-2 is presently used. In all such cases, LH-HD will achieve the same bisections at a lower latency and lower cost for Gb/s of throughput.

---
[*] Of course, there is a large number of equivalent configurations obtained via equivalence transforms N-1.

```
     1.    3F   111111           1.    1    .....1
     2.    3E   11111.           2.    2    ....1.
     3.    3D   1111.1           3.    4    ...1..
     4.    3C   1111..           4.    8    ..1...
     5.    3B   111.11           5.    10   .1....
     6.    3A   111.1.           6.    20   1.....
     7.    39   111..1           7.    7    ...111
     8.    38   111...           8.    B    ..1.11
     9.    37   11.111           9.    D    ..11.1
    10.    36   11.11.          10.    E    ..111.
    11.    35   11.1.1          11.    13   .1..11
    12.    34   11.1..          12.    15   .1.1.1
    13.    33   11..11          13.    16   .1.11.
    14.    32   11..1.          14.    19   .11..1
    15.    31   11...1          15.    1A   .11.1.
    16.    30   11....          16.    1C   .111..
    17.    2F   1.1111          17.    1F   .11111
    18.    2E   1.111.          18.    23   1...11
    19.    2D   1.11.1          19.    25   1..1.1
    20.    2C   1.11..          20.    26   1..11.
    21.    2B   1.1.11          21.    29   1.1..1
    22.    2A   1.1.1.          22.    2A   1.1.1.
    23.    29   1.1..1          23.    2C   1.11..
    24.    28   1.1...          24.    2F   1.1111
    25.    27   1..111          25.    31   11...1
    26.    26   1..11.          26.    32   11..1.
    27.    25   1..1.1          27.    34   11.1..
    28.    24   1..1..          28.    37   11.111
    29.    23   1...11          29.    38   111...
    30.    22   1...1.          30.    3B   111.11
    31.    21   1....1          31.    3D   1111.1
    32.    20   1.....          32.    3E   11111.
```

           (a)                     (b)

**Fig 4.10**

**S-2. Low Density LH networks with b=3**

This subset of LH networks is characterized by comparatively low link density and low bisection **b**=3 i.e. **B**=3$n$/2 links. They are constructed as a direct augmentation of regular hypercubic networks which have bisection **b**=1. The method is illustrated in Fig. 4.11 using augmentation of the 4-cube.

```
h₁ → 0 0 0 1
h₂ → 0 0 1 0
h₃ → 0 1 0 0
h₄ → 1 0 0 0
h₅ → 0 1 1 1
h₆ → 1 1 1 0
h₇ → 1 0 1 1
     C₁ C₂ C₃ C₄
```

**Fig. 4.11**

The $d$=4 hops $h_1$, $h_2$, $h_3$ and $h_4$ for the regular 4-cube are enclosed in a 4×4 box on the top. The augmentation consists of 3 additional hops $h_5$, $h_6$ and $h_7$ added in the form of 4 columns $C_1$, $C_2$, $C_3$ and $C_4$, where each column $C_\mu$ ($\mu$=1..$d$) has length of **L**=3 bits. The resulting network has $n$ =16 nodes with 7 links per node and it is identical to an earlier example in Fig. 4.9 with **b**=3 obtained there via translation from a $[7,4,3]_2$ EC code into the LH network. General direct construction of the **b**=3 LH network from a $d$-cube is done by appending $d$ columns $C_\mu$ ($\mu$=1..$d$) of length **L** bits, such that each bit column has *at least 2 ones* and **L** is the smallest integer satisfying inequality:

$$2^L - L - 1 \geq d \qquad (4.60)$$

The condition in eq. (4.60) expresses the requirement that $d$ columns $C_\mu$ must have at least 2 ones. Namely, there are total of $2^L$ distinct bit patterns of length **L**. Among all $2^L$ possible **L**-bit patterns, 1 pattern has 0 ones (00..0) and **L** patterns have a single one. By removing these two types, with 0 or single one, there are $2^L$-(**L**+1) remaining **L**-bit patterns with two or more ones, which is the left hand side of eq. (4.60). Any subset of $d$ distinct patterns out of these $2^L$-(**L**+1) remaining patterns can be chosen for the above augmentation. The table in Fig (4.12) shows values **L** (number of added hops to a $d$-cube) satisfying eq. (4.60) for dimensions $d$ of practical interest.

| $d_{min}$ | $d_{max}$ | L |
|---|---|---|
| 3 | 4 | 3 |
| 5 | 11 | 4 |
| 12 | 26 | 5 |
| 27 | 57 | 6 |

**Fig. 4.12**

### S-3. Augmentation of LH networks with b=odd integer

This is a very simple, yet optimal, augmentation of an LH network with $m$ links per node and bisection **b**=odd integer into LH network with bisection $b_1=b+1$ and $m_1=m+1$ links per node. The method is illustrated in Fig. 4.14 using the augmented 4-cube ($d=4$, $n=16$ nodes) with $m=7$ links per node and bisection **b**=3, which was used in earlier examples in Figures 4.9 and 4.11.

$$
\begin{array}{r|cccc}
h_1 \rightarrow & 0 & 0 & 0 & 1 \\
h_2 \rightarrow & 0 & 0 & 1 & 0 \\
h_3 \rightarrow & 0 & 1 & 0 & 0 \\
h_4 \rightarrow & 1 & 0 & 0 & 0 \\
h_5 \rightarrow & 0 & 1 & 1 & 1 \\
h_6 \rightarrow & 1 & 1 & 1 & 0 \\
h_7 \rightarrow & 1 & 0 & 1 & 1 \\
\text{XOR} & \oplus & \oplus & \oplus & \oplus \\
& \downarrow & \downarrow & \downarrow & \downarrow \\
h_8 \rightarrow & 1 & 1 & 0 & 1 \\
\end{array}
$$

**Fig. 4.14**

A single augmenting link $h_8 = h_1 \wedge h_2 \wedge \ldots \wedge h_7$ (bitwise XOR of the list) is added to the network which increases bisection from **b**=3 to **b**=4 i.e. it increases the absolute bisection **B** by $n/2=16/2=8$ links. The general augmentation method for $Cay(\mathbf{Z}_2^d, \mathbf{S}_m)$ with **b**='odd integer' consists of adding the link $h_{m+1}=h_1 \wedge h_2 \wedge \ldots \wedge h_m$ (the bitwise XOR of the previous $m$ hops) to the generator set $\mathbf{S}_m$. The resulting LH network $Cay(\mathbf{Z}_2^d, \mathbf{S}_{m+1})$ has bisection $b_1=b+1$.

The only case which requires additional computation, beyond merely XOR-ing the hop list, is the case in which the resulting hop $h_{m+1}$ happens to come out as 0 (which is an invalid hop value, a self-link of node 0 to itself). In such case, it is always possible to perform a single hop substitution in the original list $\mathbf{S}_m$ which will produce the new list with the same **b** value but a non-zero value for the list XOR result $h_{m+1}$.

## F) LH construction for a target network

In practice one would often need to construct a network satisfying requirements expressed in terms of some target number of external ports **P** having oversubscription ϕ, obtained using switches of radix **R**. The resulting construction would compute the number *n* of radix-**R** switches needed, as well as the list for detailed wiring between switches. For concreteness, each radix-**R** switch will be assumed to have **R** ports labeled as port #1, #2,… #R. Each switch will be connected to *m* other switches using ports #1, #2,… #*m* (these are topological ports or links) and leave **E** ≡ **R**-*m* ports: #*m*+1, #*m*+2,… #**R** as "external ports" per switch available to the network users for servers, routers, storage,… etc. Hence, the requirement of having total of **P** external ports is expressed in terms of **E** and number of switches *n* as:

$$\mathbf{E} = \mathbf{P}/n \qquad (4.70)$$

The oversubscription eq. (3.1) is then expressed via definition of bisection **b** in eq. (4.42) as:

$$\phi \equiv \frac{\mathbf{P}/2}{\mathbf{B}} = \frac{\mathbf{E} \cdot n/2}{\mathbf{B}} = \frac{\mathbf{E}}{\frac{\mathbf{B}}{n/2}} = \frac{\mathbf{E}}{\mathbf{b}} = \frac{\mathbf{R}-m}{\mathbf{b}} \qquad (4.71)$$

The illustrative construction below will use non-oversubscribed networks, ϕ=1, simplifying eq. (4.71):

$$\mathbf{E} = \mathbf{b} = \mathbf{R} - m \qquad (4.72)$$

i.e. for non-oversubscribed networks, the number of external ports/switch **E** must be equal to the relative bisection **b** (this the bisection in units *n*/2), or equivalently, the number of links/switch: *m* = **R** - **b**.

In order to find appropriate *n*=$2^d$ and *m* parameters, LH solutions database, obtained by translating optimum EC code tables [17] and [21] via recipe (4.45), groups solutions by network dimension *d* into record sets **D**$_d$, where *d*=3,4,… 24. These dimensions cover the range of network sizes *n*=$2^d$ that are of practical interest, from *n* = $2^3$ = 8 to *n* = $2^{24}$ ≅ 16 million switches. Each record set **D**$_d$ contains solution records for *m* = *d*, *d*+1,… $m_{max}$ links/switch, where the present database has $m_{max}$=256 links/switch. Each solution record contains, among others, the value *m*, bisection **b** and the hop list $h_1, h_2,… h_m$.

For given **P**, **R** and ϕ, LH constructor scans record sets **D**$_d$, for *d*=3,4,… and in each set, inspects the records for *m*=*d*, *d*+1, … computing for each (*d,m*) record values **E**(*d,m*)=**R**-*m* ports/switch, total ports **P**(*d,m*) = *n*· **E**(*d,m*) = $2^d$·(**R**-*m*) and oversubscription ϕ(*d,m*)=**E**(*d,m*)/**b** (value **b** is in each (*d,m*) record). The relative errors δ**P** = |**P**(*d,m*)-**P**|/**P** and δϕ = |ϕ(*d,m*)- ϕ|/ϕ are computed and the best match (record (*d,m*) with the lowest combined error) is selected as the solution to use. If the requirement is "at least **P** ports" then the constraint **P**(*d,m*)-**P**≥0 is imposed for the admissible comparisons. The requirements can also prioritize δ**P** and δϕ via weights for each (e.g. 0.7·δ**P** + 0.3·δϕ for total error). After finding the best matching (*d,m*) record, the hop list $h_1, h_2,… h_m$ is retrieved from the record and the set of links **L**(*v*) is computed for each node *v*, where *v* = 0, 1, … *n*-1, as: **L**(*v*) = { *v*^$h_s$ for *s*=1..*m*}. Given *n* such sets of links, **L**(0), **L**(1),..., **L**(*n*-1), the complete wiring for the network is specified. The examples below illustrate the described construction procedure.

**Example 1.** Small network with **P**=96 ports at $\phi$=1, using switches with radix **R**=12

The LH database search finds the exact match (δ**P**=0, δ$\phi$=0) for the record **d**=5, **m**=9, hence requiring **n**=$2^d$=$2^5$=32 switches of radix **R**=12. The bisection **b**=3 and the hop list (in hex base) for the record is: $S_9$={1, 2, 4, 8, 10, E, F, 14, 19}$_{hex}$. The number of external ports per switch is **E**=**b**=3, combined with **m**=9 topological ports/switch, results in radix **R**=3+9=12 total ports/switch as specified. The total number of external ports is **P** = **E**·**n** = 3·32 = 96 as required. Diameter (max hops) for the network is **D**=3 hops, and the average hops (latency) is **Avg**=1.6875 hops. The table in Fig. 4.15 shows complete connection map for the network for 32 switches, stacked in a 32-row rack one below the other, labeled in leftmost column "Sw" as 0, 1,… 1F (in hex). Switch 5 is outlined with connections shown for its ports #1,#2,… #9 to switches (in hex) 04, 07, 01, 0D, 15, 0B, 0A, 11 and 1C. These 9 numbers are computed by XOR-ing 5 with the 9 generators (row 0): 01, 02, 04, 08, 10, 0E, 0F, 14, 19. The free ports are #10, #11 and #12.

| Sw/Pt: | #1 | #2 | #3 | #4 | #5 | #6 | #7 | #8 | #9 | #10 | #11 | #12 |
|---|---|---|---|---|---|---|---|---|---|---|---|---|
| 0:  | 01 | 02 | 04 | 08 | 10 | 0E | 0F | 14 | 19 | ** | ** | ** |
| 1:  | 00 | 03 | 05 | 09 | 11 | 0F | 0E | 15 | 18 | ** | ** | ** |
| 2:  | 03 | 00 | 06 | 0A | 12 | 0C | 0D | 16 | 1B | ** | ** | ** |
| 3:  | 02 | 01 | 07 | 0B | 13 | 0D | 0C | 17 | 1A | ** | ** | ** |
| 4:  | 05 | 06 | 00 | 0C | 14 | 0A | 0B | 10 | 1D | ** | ** | ** |
| 5:  | 04 | 07 | 01 | 0D | 15 | 0B | 0A | 11 | 1C | ** | ** | ** |
| 6:  | 07 | 04 | 02 | 0E | 16 | 08 | 09 | 12 | 1F | ** | ** | ** |
| 7:  | 06 | 05 | 03 | 0F | 17 | 09 | 08 | 13 | 1E | ** | ** | ** |
| 8:  | 09 | 0A | 0C | 00 | 18 | 06 | 07 | 1C | 11 | ** | ** | ** |
| 9:  | 08 | 0B | 0D | 01 | 19 | 07 | 06 | 1D | 10 | ** | ** | ** |
| A:  | 0B | 08 | 0E | 02 | 1A | 04 | 05 | 1E | 13 | ** | ** | ** |
| B:  | 0A | 09 | 0F | 03 | 1B | 05 | 04 | 1F | 12 | ** | ** | ** |
| C:  | 0D | 0E | 08 | 04 | 1C | 02 | 03 | 18 | 15 | ** | ** | ** |
| D:  | 0C | 0F | 09 | 05 | 1D | 03 | 02 | 19 | 14 | ** | ** | ** |
| E:  | 0F | 0C | 0A | 06 | 1E | 00 | 01 | 1A | 17 | ** | ** | ** |
| F:  | 0E | 0D | 0B | 07 | 1F | 01 | 00 | 1B | 16 | ** | ** | ** |
| 10: | 11 | 12 | 14 | 18 | 00 | 1E | 1F | 04 | 09 | ** | ** | ** |
| 11: | 10 | 13 | 15 | 19 | 01 | 1F | 1E | 05 | 08 | ** | ** | ** |
| 12: | 13 | 10 | 16 | 1A | 02 | 1C | 1D | 06 | 0B | ** | ** | ** |
| 13: | 12 | 11 | 17 | 1B | 03 | 1D | 1C | 07 | 0A | ** | ** | ** |
| 14: | 15 | 16 | 10 | 1C | 04 | 1A | 1B | 00 | 0D | ** | ** | ** |
| 15: | 14 | 17 | 11 | 1D | 05 | 1B | 1A | 01 | 0C | ** | ** | ** |
| 16: | 17 | 14 | 12 | 1E | 06 | 18 | 19 | 02 | 0F | ** | ** | ** |
| 17: | 16 | 15 | 13 | 1F | 07 | 19 | 18 | 03 | 0E | ** | ** | ** |
| 18: | 19 | 1A | 1C | 10 | 08 | 16 | 17 | 0C | 01 | ** | ** | ** |
| 19: | 18 | 1B | 1D | 11 | 09 | 17 | 16 | 0D | 00 | ** | ** | ** |
| 1A: | 1B | 18 | 1E | 12 | 0A | 14 | 15 | 0E | 03 | ** | ** | ** |
| 1B: | 1A | 19 | 1F | 13 | 0B | 15 | 14 | 0F | 02 | ** | ** | ** |
| 1C: | 1D | 1E | 18 | 14 | 0C | 12 | 13 | 08 | 05 | ** | ** | ** |
| 1D: | 1C | 1F | 19 | 15 | 0D | 13 | 12 | 09 | 04 | ** | ** | ** |
| 1E: | 1F | 1C | 1A | 16 | 0E | 10 | 11 | 0A | 07 | ** | ** | ** |
| 1F: | 1E | 1D | 1B | 17 | 0F | 11 | 10 | 0B | 06 | ** | ** | ** |

**Fig 4.15**

To illustrate the interpretation of the links via numbers, the outlined switch "5:" indicates on its port #2 a connection to switch 7 (the encircled number 07 in the row 5:). In the row 7:, labeled as switch "7:", there is an encircled number 05 at its port #2 (column #2), which refers back to this same connection between the switch 5 and the switch 7 via port #2 on each switch. The same pattern can be observed between any pair of connected switches and ports.

**Example 2.** Small network with **P**=1536 (1.5K) ports at $\phi$=1, using switches with radix **R**=24.

The LH solutions database search finds an exact match for $d = 8$, $n = 256$ switches of radix **R**=24 and $m$=18 topological ports/switch. Diameter (max hops) of the network is **D**=3 hops, and average latency is **Avg**=2.2851562 hops. The bisection is **b**=6, providing thus **E**=6 free ports per switch at $\phi$=1. The total number of ports provided is **E**·$n$=6·256=1536 as required. The set of 18 generators is: $S_{18}$ = { 01, 02, 04, 08, 10, 20, 40, 80, 1A, 2D, 47, 78, 7E, 8E, 9D, B2, D1, FB}$_{hex}$. Note that the first 8 links are regular 8-cube links (power of 2), while the remaining 10 are LH augmentation links. These generators specify the target switches (as index 00..FF$_{hex}$) connected to switch 00 via ports #1, #2,… #18 (switches on both ends of a link use the same port number for mutual connections). To compute the 18 links (to 18 target switches) for some other switch $x \neq$ 00, one would simply XOR number $x$ with the 18 generators. Fig. 4.16 shows the connection table only for the first 16 switches of the resulting network, illustrating this computation of the links. For example, switch 1 (row '1:') has on its port #4 target switch 09, which is computed as 1^8=9, where 8 was the generator in row '0:' for port #4. Checking then switch 9 (in row '9:'), on its port #4 is switch 01 (since 9^8=1), i.e. switches 1 and 9 are connected via port #4 on each. The table also shows that each switch has 6 ports #19, #20,… #24 free.

```
Sw/Pt:  #1  #2  #3  #4  #5  #6  #7  #8  #9  #10 #11 #12 #13 #14 #15 #16 #17 #18 #19 #20 #21 #22 #23 #24
   0:   01  02  04  08  10  20  40  80  1A  2D  47  78  7E  8E  9D  B2  D1  FB  **  **  **  **  **  **
   1:   00  03  05 (09) 11  21  41  81  1B  2C  46  79  7F  8F  9C  B3  D0  FA  **  **  **  **  **  **
   2:   03  00  06  0A  12  22  42  82  18  2F  45  7A  7C  8C  9F  B0  D3  F9  **  **  **  **  **  **
   3:   02  01  07  0B  13  23  43  83  19  2E  44  7B  7D  8D  9E  B1  D2  F8  **  **  **  **  **  **
   4:   05  06  00  0C  14  24  44  84  1E  29  43  7C  7A  8A  99  B6  D5  FF  **  **  **  **  **  **
   5:   04  07  01  0D  15  25  45  85  1F  28  42  7D  7B  8B  98  B7  D4  FE  **  **  **  **  **  **
   6:   07  04  02  0E  16  26  46  86  1C  2B  41  7E  78  88  9B  B4  D7  FD  **  **  **  **  **  **
   7:   06  05  03  0F  17  27  47  87  1D  2A  40  7F  79  89  9A  B5  D6  FC  **  **  **  **  **  **
   8:   09  0A  0C  00  18  28  48  88  12  25  4F  70  76  86  95  BA  D9  F3  **  **  **  **  **  **
   9:   08  0B  0D (01) 19  29  49  89  13  24  4E  71  77  87  94  BB  D8  F2  **  **  **  **  **  **
   A:   0B  08  0E  02  1A  2A  4A  8A  10  27  4D  72  74  84  97  B8  DB  F1  **  **  **  **  **  **
   B:   0A  09  0F  03  1B  2B  4B  8B  11  26  4C  73  75  85  96  B9  DA  F0  **  **  **  **  **  **
   C:   0D  0E  08  04  1C  2C  4C  8C  16  21  4B  74  72  82  91  BE  DD  F7  **  **  **  **  **  **
   D:   0C  0F  09  05  1D  2D  4D  8D  17  20  4A  75  73  83  90  BF  DC  F6  **  **  **  **  **  **
   E:   0F  0C  0A  06  1E  2E  4E  8E  14  23  49  76  70  80  93  BC  DF  F5  **  **  **  **  **  **
   F:   0E  0D  0B  07  1F  2F  4F  8F  15  22  48  77  71  81  92  BD  DE  F4  **  **  **  **  **  **
  10:   …
```

**Fig. 4.16**

**Example 3.** Large network with **P**=655,360 (640K) ports at $\phi$=1, using switches with radix **R**=48.

The database lookup finds the exact match using $d$=16, $n$=$2^{16}$ = 65,536 = 64K switches of radix **R**=48. Each switch uses $m$=38 ports for connections with other switches leaving **E**=48-38=10 ports/switch free, yielding total of $\mathbf{P} = \mathbf{E} \cdot \mathbf{n} = 10 \cdot 64K$=640K available ports. Bisection is **b**=10 resulting in $\phi$=E/b=1. The list of $m$=38 generators $S_{38}$ = {$h_1, h_2, \ldots h_{38}$} is shown in Fig. 4.17 in hex and binary base. The 38 links for some switch $x$ (where $x$: 0..FFFF) are computed as $S_{38}(x) \equiv \{x \wedge h_1, x \wedge h_2, \ldots x \wedge h_{38}\}$. Diameter (max hops) of the network is **D**=5 hops, and the average latency is **Avg**=4.061691 hops.

```
 1.     1    ..............1
 2.     2    .............1.
 3.     4    ............1..
 4.     8    ...........1...
 5.    10    ..........1....
 6.    20    .........1.....
 7.    40    ........1......
 8.    80    .......1.......
 9.   100    ......1........
10.   200    .....1.........
11.   400    ....1..........
12.   800    ...1...........
13.  1000    ...1...........
14.  2000    ..1............
15.  4000    .1.............
16.  8000    1..............
17.   6F2    .....11.1111..1.
18.  1BD0    ...11.1111.1....
19.  1F3D    ...11111..1111.1
20.  3D72    ..1111.1.111..1.
21.  6B64    .11.1.11.11..1..
22.  775C    .111.111.1.111..
23.  893A    1...1..1..111.1.
24.  8B81    1...1.111......1
25.  9914    1..11..1...1.1..
26.  A4C2    1.1..1..11....1.
27.  A750    1.1..111.1.1....
28.  B70E    1.11.111....111.
29.  BFF1    1.1111111111...1
30.  C57D    11...1.1.11111.1
31.  D0A6    11.1....1.1..11.
32.  D1CA    11.1...111..1.1.
33.  E6B5    111..11.1.11.1.1
34.  EAB9    111.1.1.1.111..1
35.  F2E8    1111..1.111.1...
36.  F313    1111..11...1..11
37.  F9BF    11111..11.111111
38.  FC31    111111....11...1
```

**Fig. 4.17**

## G) LH performance comparisons

The LH solutions database (containing ~3300 LH configurations) was used to compare LH networks against several leading alternatives from industry and research across broader spectrum of parameters. The comparison charts are shown in figures 4.20-4.24. The metrics used for evaluation were **Ports/Switch** yield (ratio **P/n**, higher is better) and the cables consumption as **Cables/Port** (ratio: # of topological cables/**P**, lower is better). In order to maximize the fairness of the comparisons, the alternative networks were set up to generate some number of ports **P** using switches of radix **R**, which are optimal parameters values for a given alternative network (each network type has its own "natural" parameter values at which it produces the most efficient networks). Only then the LH network was constructed to match the given number of external ports **P** using switches of radix **R** (as a rule, these are not the optimal or "natural" parameters for LH networks). Full details of these computations, including derivations of performance formulas for all alternative topologies is available in a separate tech note TN12-0108.pdf and in a spreadsheet LHCalc.xlsx used to compute the charts shown.

The chart for each alternative network shows Ports/Switch yields for the LH network ‒ ‒ ‒ and the alternative network ....., along with the ratio LH/alternative ⎯⎯ with numbers on the right axis (e.g. a ratio 3 means that LH yields 3 times more Ports/Switch than the alternative). The second chart for each alternative network shows the Cables/Port consumption for the LH and the alternative, along with the ratio: alternative/LH on the right axis (e.g. a ratio 3 means that LH consumes 3 times fewer cables per port produced than the alternative). All networks are non-oversubscribed i.e. $\phi=1$.

### 1) LH vs. Hypercube

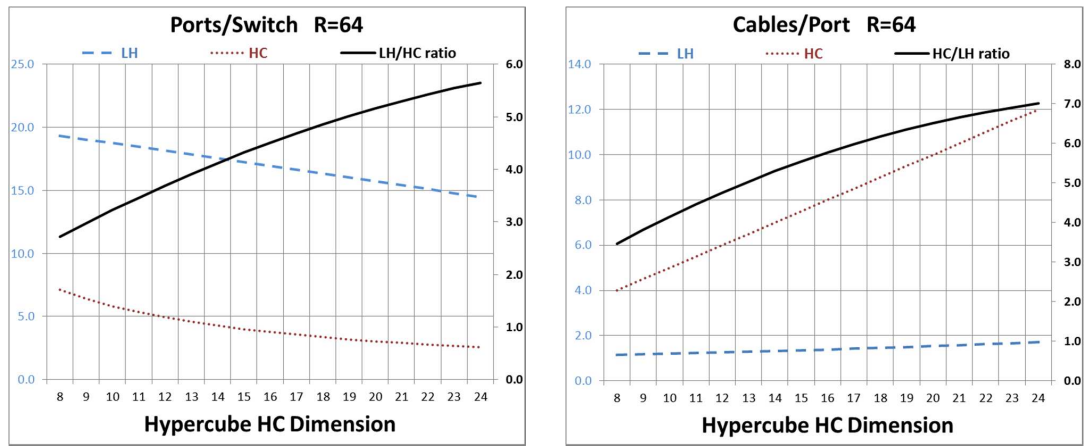

**Fig. 4.20**

For example, the Ports/Switch chart shows yield for hypercube (HC), for network sizes from $n=2^8$ to $2^{24}$ switches of radix **R**=64. The Ports/Switch for LH network yielding the same total number of ports **P** is shown, along with the ratio LH/HC, which shows (on the right axis scale) that LH produces 2.6 to 5.8 times greater Ports/Switch yield than hypercube, hence it uses 2.6-5.8 times fewer switches than HC to produce the same number of ports **P** as HC at the same throughput. The second chart shows similarly the Cables/Port consumption for HC and LH, and the ratio HC/LH of the two (right axis scale), showing that LH consumes 3.5 to 7 times fewer cables to produce the same number of ports **P** as HC at the same throughput. The remaining charts show the same type of comparisons for the other four alternatives.

**2) LH vs. Folded Cube (FC)**

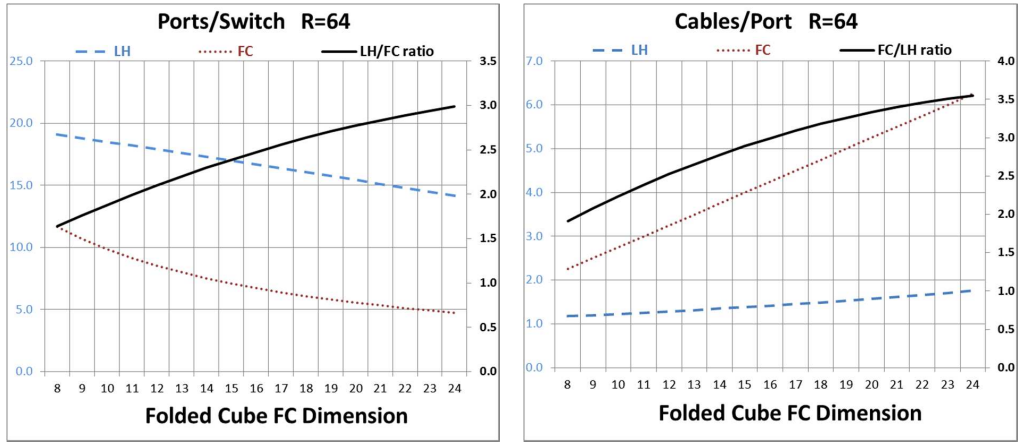

**Fig. 4. 21**

**3) LH vs. Flattened Butterfly (FB)**

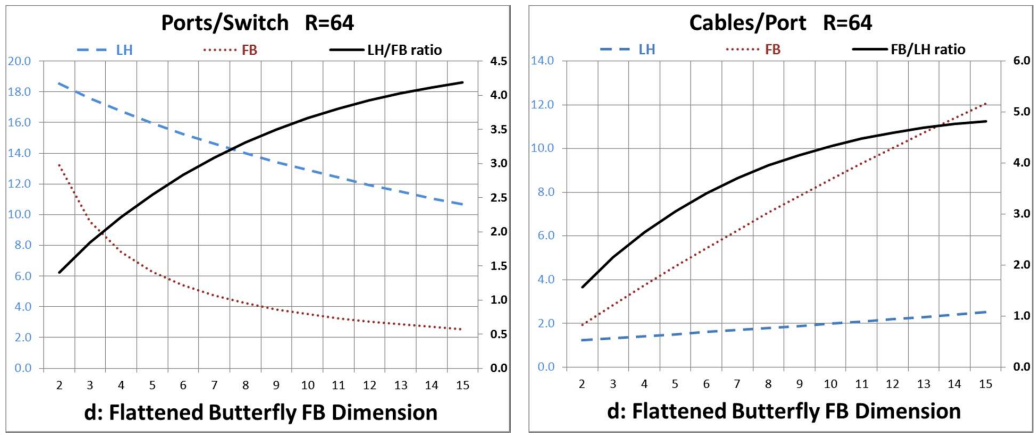

**Fig. 4. 22**

## 4) LH vs. Fat Tree (FT)

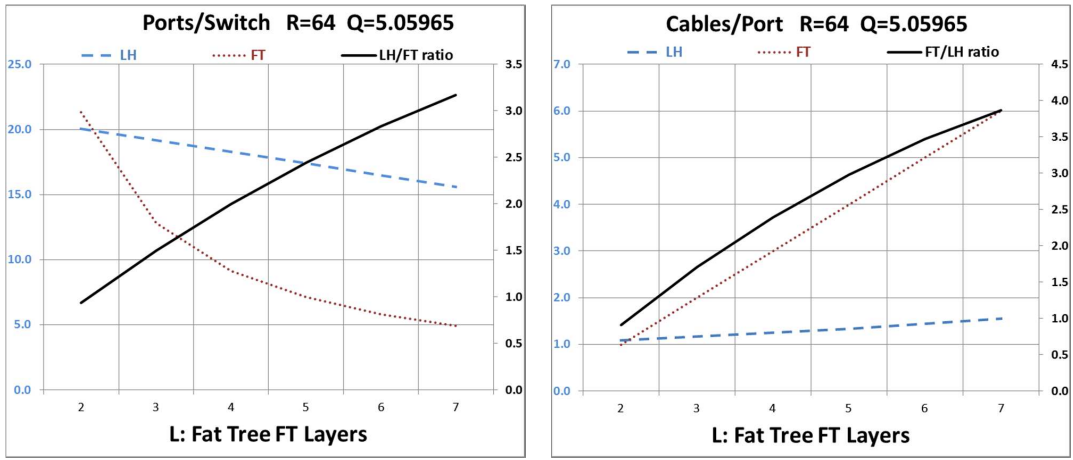

**Fig. 4. 23**

## 5) LH vs. Dragonfly (DF)

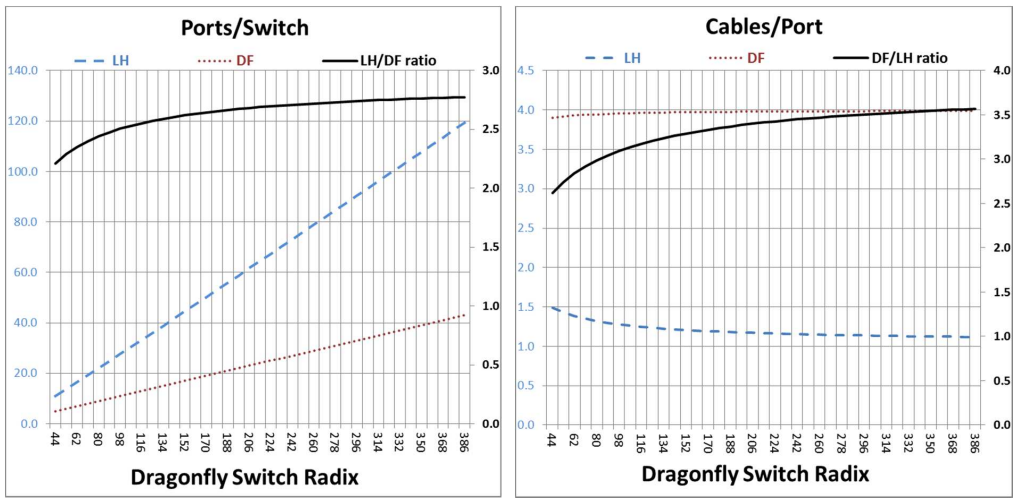

**Fig. 4. 24**